\documentclass[10pt]{article}
\usepackage[margin=1in,nohead,centering]{geometry}
\usepackage{amsmath,amssymb}

\usepackage{caption}

\usepackage{cite}

\usepackage[table]{xcolor}

\usepackage{array}

\usepackage{graphicx}
\usepackage{epstopdf}

\usepackage{graphicx,color,wrapfig}
\usepackage{amssymb,amsmath}
\usepackage{url}
\usepackage{xspace}
\usepackage{float}
\usepackage{graphics}
\usepackage{soul}
\usepackage{subcaption}
\usepackage{textcomp} 
\usepackage{listings} 

\newcommand{\aseq}{{\bf a}\xspace}

\newtheorem{theorem}{Theorem}

\newcommand{\op}{\,\mbox{\bf\texttt{(}}\,}
\newcommand{\cp}{\,\mbox{\bf\texttt{)}}\,}

\title{RNAmountAlign: efficient software for local, global, semiglobal pairwise and multiple RNA sequence/structure alignment} 
\author{Amir H. Bayegan \and Peter Clote\thanks{Corresponding author: 
{\tt clote@bc.edu}}}
\date{Biology Department, Boston College, Chestnut Hill, MA 02467, USA}

\begin{document}
\maketitle
\begin{abstract}
Alignment of structural RNAs is an important problem with a wide range of applications. Since function is often determined by molecular structure, RNA alignment programs should take into account both sequence and base-pairing information for structural homology identification. A number of successful alignment programs are heuristic versions of Sankoff's optimal algorithm. Most of them require $O(n^4)$ run time. This paper describes {\tt C++} software, {\tt RNAmountAlign}, for RNA sequence/structure alignment that runs in $O(n^3)$ time and $O(n^2)$ space; moreover, our software returns a $p$-value (transformable to expect value $E$) based on Karlin-Altschul statistics for local alignment, as well as parameter fitting for local and global alignment. 
Using incremental mountain height, a representation of structural information computable in cubic time, {\tt RNAmountAlign} implements quadratic time pairwise local, global and global/semiglobal (query search) alignment using a weighted combination of sequence and structural similarity. {\tt RNAmountAlign} is capable of performing progressive multiple alignment as well. Benchmarking of
{\tt RNAmountAlign} against {\tt LocARNA}, {\tt LARA}, {\tt FOLDALIGN}, {\tt DYNALIGN} and {\tt STRAL} shows that {\tt RNAmountAlign} has reasonably good
accuracy and much faster run time supporting all alignment types.
\\
\noindent
\textbf{Availability:}  
{\tt RNAmountAlign} is publicly available at
\url{http://bioinformatics.bc.edu/clotelab/RNAmountAlign}.
\end{abstract}

\section{Introduction}
A number of different metrics exist for comparison of RNA secondary structures, including 
base pair distance (BP), string edit distance (SE) \cite{levenshteinDistance},
mountain distance (MD) \cite{moulton},
tree edit distance (TE) \cite{Shapiro.cab88},
coarse tree edit distance (HTE) \cite{Lorenz.amb11}, morphological distance \cite{Voss.b04} and a few other metrics.
In what appears to be the most comprehensive published
comparison of various secondary structure metrics \cite{mountainHeight},
it was shown that all of these distance measures are highly correlated
when computing distances between structures taken from the Boltzmann 
low-energy ensemble of secondary structures \cite{dingLawrence} 
for the same RNA sequence -- so-called {\em intra-ensemble} correlation.
In contrast,  these distance measures have low correlation
when computing distances between structures taken from Boltzmann 
ensembles of different RNA sequences of the same length -- so-called
{\em inter-ensemble} correlation. 
For instance, the intra-ensemble correlation between base pair distance
(BP) and mountain distance (MD) is $0.822$, while the corresponding
inter-ensemble correlation drops to $0.210$.  Intra-ensemble correlation 
between string edit distance (SE) and the computationally more expensive 
tree edit distance (TE) is $0.975$, while the corresponding intra-ensemble
correlation drops to $0.590$ -- see Table~\ref{table:correlationMetrics}.

\begin{table}[]
\centering
\begin{tabular}{|c|ccccc|}
\hline
    & BP    & MD    & SE    & TE    & HTE   \\ \hline
BP  &       & 0.210 & 0.134 & 0.133 & 0.230 \\
MD  & 0.822 &       & 0.519 & 0.607 & 0.515 \\
SE  & 0.960 & 0.853 &       & 0.590 & 0.310 \\
TE  & 0.943 & 0.879 & 0.975 &       & 0.597 \\
HTE & 0.852 & 0.844 & 0.879 & 0.913 &       \\ \hline
\end{tabular}
\caption{Correlation between various secondary structure metrics, as
computed in \cite{mountainHeight}:
base pair distance (BP), string edit distance (SE) \cite{levenshteinDistance},
mountain distance (MD) \cite{moulton},
tree edit distance (TE) \cite{Shapiro.cab88} and
coarse tree edit distance (HTE) \cite{Lorenz.amb11}.
Lower triangular values indicate intra-ensemble correlations; upper triangular
values indicate inter-ensemble correlations. Table values are taken from \cite{mountainHeight}.
}
\label{table:correlationMetrics}
\end{table}
\medskip

Due to poor inter-ensemble correlation of RNA secondary structure metrics,
and the fact that most secondary structure pairwise alignment algorithms
depend essentially on some form of base pair distance, string edit
distance, or free energy of common secondary structure, we have developed 
the first RNA sequence/structure pairwise
alignment algorithm that is based on (incremental ensemble) 
mountain distance. Our software, {\tt RNAmountAlign}, uses this distance
measure, since the Boltzmann ensemble of all secondary structures of a 
given RNA of length $n$ can represented as a length $n$ vector of real
numbers, thus allowing an adaptation of fast sequence alignment methods.
Depending on the command-line flag given, our software, {\tt RNAmountAlign}
can perform pairwise alignment, 
(Needleman-Wunsch global \cite{Needleman.jmb70}, 
Smith-Waterman local \cite{smithWaterman} or 
semiglobal \cite{gusfield} alignment) as well as
progressive multiple alignment (global and local),
computed using a guide tree as in {\tt CLUSTAL} \cite{Thompson.nar99}.
Expect values $E$ for local alignments are computed using
Karlin-Altschul extreme-value statistics \cite{Karlin.pnas90,karlinDemboKawabata}, suitably
modified to account for
our new sequence/structure similarity measure.
Additionally, {\tt RNAmountAlign} can determine $p$-values (hence $E$-values)
by parameter fitting for the normal (ND), extreme value (EVD) and
gamma (GD) distributions.

We benchmark the performance of {\tt RNAmountAlign}
on pairwise and multiple global sequence/structure alignment of RNAs 
against the widely used programs {\tt LARA},
{\tt FOLDALIGN}, {\tt DYNALIGN}, {\tt LocARNA} and {\tt STRAL}.
{\tt LARA} (Lagrangian relaxed structural alignment) \cite{Bauer.bb07} formulates the problem of RNA (multiple) sequence/structure 
alignment as a problem in integer linear programming (ILP), then computes
optimal or near-optimal solutions to this problem. The software 
{\tt FOLDALIGN} \cite{foldalign,Havgaard.cpb12,Sundfeld.b16}, and 
{\tt DYNALIGN} \cite{Mathews.jmb02} are different $O(n^4)$ approximate
implementations of Sankoff's $O(n^6)$ optimal RNA sequence/structure 
alignment algorithm. {\tt FOLDALIGN} sets limits on the maximum length 
of the alignment as well as the maximum distance between subsequences 
being aligned in order to reduce the time complexity of the Sankoff algorithm. 
{\tt DYNALIGN} \cite{Mathews.jmb02}
implements pairwise RNA secondary structural alignment by
determining the common structure to both sequences that has lowest free
energy, using a positive (destabilizing) energy heuristic for gaps
introduced, in addition to setting bounds on the distance between subsequences
being aligned. In particular, the only contribution from nucleotide 
information in {\tt Dynalign} is from the nucleotide-dependent
free energy parameters for base stacking, dangles, etc. 
{\tt LocARNA} (local alignment of RNA)  \cite{Will.pcb07,Smith.nar10} 
is a heuristic implementation of {\tt PMcomp} \cite{Hofacker2004} which 
compares the base pairing probability matrices computed by 
McCaskill's algorithm. 
Although the software is not maintained, {\tt STRAL} \cite{Dalli.b06} which is similar to our approach, uses up- and downstream base pairing 
probabilities as the structural information and combines them with 
sequence similarity in a weighted fashion.

{\tt LARA}, {\tt mLocARNA} (extension of {\tt LocARNA}), 
{\tt FOLDALIGNM} \cite{Torarinsson.b07,Havgaard.cpb12}
(extension of {\tt FOLDALIGN}), 
{\tt Multilign} \cite{Xu.b11,Xu.mmb16} 
(extension of {\tt DYNALIGN}) and {\tt STRAL}
support multiple alignment. 
{\tt LARA} computes all pairwise sequence alignments and 
subsequently uses the {\tt T-Coffee} package \cite{NOTREDAME2000} 
to construct multiple alignments. 
Both {\tt FOLDALIGNM} and {\tt mLocARNA} implement progressive 
alignment of consensus base pairing probability matrices using a 
guide tree similar to the approach of {\tt PMmulti} \cite{Hofacker2004}. 
For a set of given sequences, {\tt Multilign} uses {\tt DYNALIGN} to 
compute the pairwise alignment of a single fixed index sequence to 
each other sequence in the set, and computes a consensus structure. 
In each pairwise alignment, only the index sequence base pairs found in 
previous computations are used. More iterations in the same manner 
with the same index sequence are then used to improve the structure 
prediction of other sequences. The number of pairwise alignments in 
{\tt Multilign} is linear with respect to the number of sequences. 
{\tt STRAL} performs multiple alignment in a fashion similar to 
{\tt CLASTALW} \cite{Thompson1994}. 
Table~\ref{table:softwareOverview} provides an overview of various
features, to the best of our knowledge, supported by the software 
benchmarked in this paper.

\begin{table}
\centering
\begin{small}
\begin{tabular}{|l|c c c c c c|}
\hline									
Software	&	Local	&	Global	&	Semiglobal	&	E-value	& F1(Pairwise) & SPS(Multiple)\\
\hline									
RNAmountAlign	&	$\checkmark$	&	$\checkmark$	&	$\checkmark$	&	$\checkmark$	& 0.84 & 0.84 \\
LocARNA	&	$\checkmark$	&	$\checkmark$	&	\textemdash	&	\textemdash	& 0.81 & 0.84\\
LARA	&	\textemdash	&	$\checkmark$	&	\textemdash	&	\textemdash	& 0.84 & 0.85 \\
FOLDALIGN	&	$\checkmark$	&	$\checkmark$	&	\textemdash	&	$\checkmark$	& 0.80 & 0.77 \\
DYNALIGN	&	\textemdash	&	$\checkmark$	&	\textemdash	&	\textemdash	& 0.68 & 0.67 \\
STRAL	&	\textemdash	&	$\checkmark$	&	\textemdash	&	\textemdash	& 0.82 & -\\
\hline									
\end{tabular}
\end{small}
\medskip
\caption{Overview of features in software used in benchmarking tests,
where $\checkmark$ [resp. \textemdash] indicates the presence [resp.
absence] of said feature, to the best of our knowledge. Average F1 [resp. SPS] scores for the pairwise [resp. multiple] global alignment are given, computed as explained in the text.
}
\label{table:softwareOverview}
\end{table}

{\tt RNAmountAlign} can perform semiglobal alignments in addition to global and local alignments. As in the RNA tertiary structural alignment software {\tt DIAL}
\cite{Ferre.nar07}, semiglobal alignment allows the user to perform a query 
search, where the query is entirely matched to a local portion of the target.
Quadratic time alignment using affine gap cost is implemented in
{\tt RNAmountAlign} using the Gotoh method \cite{Gotoh.jmb82} with 
the following pseudocode, shown for the case of semiglobal alignment.  
Let $g(k)$ denote an affine cost for size $k$ gap, defined by $g(0)=0$ and 
$g(k) = g_i  + (k-1)\cdot g_e$ for positive gap initiation [resp. extension]
costs $g_i$ [resp. $g_e$].
For query ${\bf a} = a_1,\ldots,a_n$ and target ${\bf b} = b_1,\ldots,b_m$, 
define $(n+1)$ $\times$ $(m+1)$ matrices $M,P,Q$ as follows: 
$M_{i,0}=g(i)$ for all $1 \leq i \leq n$,
$M_{0,j}=0$ for all $1\leq j \leq m$, while for positive $i,j$ we have
$M_{i,j} =$ $\max$ $( M_{i-1,j-1} +$
$\mbox{sim}$ $(a_i,b_j)$, $P_{i,j}$, $Q_{i,j})$.
For $1 \leq i \leq n$, $1 \leq j \leq m$, let
$P_{0,j} = 0$ and
$P_{i,j} = \max \left( M_{i-1,j} + g_i, P_{i-1,j} + g_e \right)$,
and define
$Q_{i,0} = 0$ and
$Q_{i,j} = \max\left( M_{i,j-1} + g_i, Q_{i,j-1} + g_e, 0 \right)$.
Determine the maximum semiglobal alignment score in row $n$, then
perform backtracking to obtain an optimal semiglobal (or query search)
alignment.\\

In this paper we provide a very fast, comprehensive software package capable of pairwise/multiple local/global/semiglobal alignment with $p$-values and $E$-values for statistical significance. Moreover, due to its speed and relatively good accuracy, the software can be used for whole-genome searches for homologues of a given orphan RNA as query. This is in contrast to {\tt Infernal} \cite{eddy.bioinf2013}, which requires a multiple alignment to construct a covariance model for whole-genome searches.
\section{Materials and methods}

\subsubsection{Incremental ensemble expected mountain height}

Introduced in \cite{Hogeweg.nar84},
the {\em mountain height}\footnote{We follow \cite{Hogeweg.nar84,moulton}
in our definition of mountain height, and related notions of ensemble
mountain height and distance, while \cite{Huynen1996} 
and Vienna RNA package \cite{Lorenz.amb11} 
differ in an inessential manner by defining
$h_s(k) = |\{ (i,j) \in s : i < k \}| - |\{ (i,j) \in s : j \leq k \}|$.}
$h_s(k)$ of secondary structure $s$ at position 
$k$ is defined as the number of base pairs in $s$ that lie between an
external loop and $k$, formally given by
\begin{align}
\label{eqn:mountainHeight}
h_s(k) &= 
|\{ (i,j) \in s : i \leq k \}| -
|\{ (i,j) \in s : j \leq k \}| 
\end{align}

\begin{figure}
\centering
\includegraphics[width=0.6\textwidth]{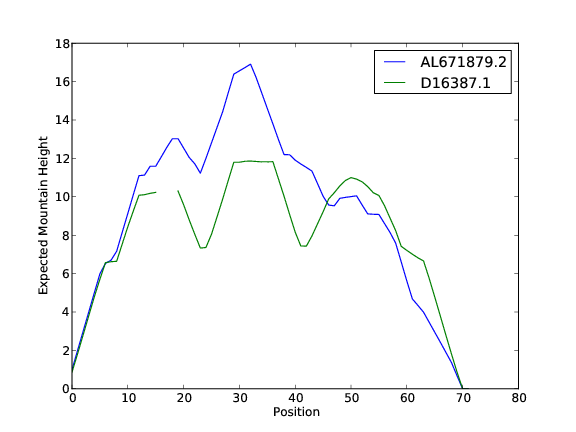}
\caption{Ensemble mountain heights of
72 nt tRNA AL671879.2 and 69 nt tRNA D16387.1, aligned together by
{\tt RNAmountAlign}. Since the
{\tt BRAliBase 2.1 K2} reference (pairwise) alignment \cite{Gardner.nar05} 
has only 28\% sequence identity, 
structural similarity parameter $\gamma$ was set to $1$ in
our software {\tt RNAmountAlign}, which returned the correct
alignment. See Methods section for explanation of $\gamma$ and the
algorithm used by {\tt RNAmountAlign}.
}
\label{fig:alignmentTRNAwith28percentSeqId}
\end{figure}

The {\em ensemble mountain height} $\langle h(k) \rangle$ 
\cite{Huynen1996} for RNA sequence 
${\bf a}=a_1,\ldots,a_n$ at position $k$ is defined as the average
mountain height, where the average is taken over the Boltzmann ensemble
of all low-energy structures $s$ of sequence ${\bf a}$. 
If base pairing probabilities $p_{i,j}$ have been
computed, then it follows that 
\begin{align}
\label{eqn:ensembleMountainHeight}
\langle h(k) \rangle &= \sum\limits_{i\leq k} p_{i,j}
- \sum\limits_{j \leq k} p_{i,j}
\end{align}
and hence the {\em incremental ensemble mountain height}, which for
values $1<k\leq n$ is defined by
$m_{\bf a}(k) = \langle h(k) \rangle - \langle h(k-1) \rangle$ 
can be readily computed by
\begin{align}
\label{eqn:incrementalEnsembleMountainHeight}
m_{\bf a}(k) &= \left\{ \begin{array}{ll}
0 &\mbox{if $k=1$}\\
\sum\limits_{k < j} p_{k,j} - \sum\limits_{i < k} p_{i,k}
&\mbox{else} 
\end{array} \right.
\end{align}
It is clear that $-1 \leq m_{\bf a}(k) \leq 1$, and that both
ensemble mountain height and incremental ensemble mountain height can be
computed in time that is quadratic in sequence length $n$, provided that
base pairing probabilities $p_{i,j}$ have been computed. Except for the
cubic time taken by a function call of {\tt RNAfold} from Vienna RNA
package \cite{Lorenz.amb11}, the software {\tt RNAmountAlign} has 
quadratic time and space requirements. 
Figure~\ref{fig:alignmentTRNAwith28percentSeqId} depicts a global alignment
of two transfer RNAs, computed by {\tt RNAmountAlign}, shown as 
superimposed ensemble mountain height displays with gaps.

\subsubsection{Transforming distance into similarity}

In \cite{smithWatermanSellers},
Seller's (distance-based) global pairwise alignment algorithm 
\cite{sellersAlignmentDistance}
was rigorously shown to be equivalent to 
Needleman and Wunsch's (similarity-based) global pairwise alignment 
algorithm \cite{Needleman.jmb70}.  Recalling that Seller's alignment distance 
is defined as the minimum, taken over all alignments of the sum
of distances $d(x,y)$ between aligned nucleotides
$x,y$ plus the sum of (positive) weights $w(k)$ for size $k$ gaps, 
while Needleman-Wunsch alignment similarity is defined as the maximum,
taken over all alignments of the sum of similarities $s(x,y)$ between 
aligned nucleotides $x,y$ plus the sum of
(negative) gap weights $g(k)$ for size $k$ gaps, 
Smith and Waterman \cite{smithWatermanSellers} show that by defining
\begin{eqnarray}
\label{eqn:distanceFromSimilarity}
d(x,y) &= \max\limits_{\scriptsize a,b \in \{A,C,G,U\}} s(a,b) - s(x,y)\\
\label{eqn:distanceGapFromSimilarityGap}
w(k) &= \frac{k}{2} \cdot 
\max\limits_{\scriptsize a,b \in \{A,C,G,U\}} s(a,b) - g(k)
\end{eqnarray}
and by taking the minimum distance, rather than maximum similarity,
the Needleman-Wunsch algorithm is transformed into Seller's algorithm.
Though formulated here for RNA nucleotides, equivalence holds over 
arbitrary alphabets and similarity measures (e.g.  BLOSUM62).

For $x,y \in \{ \op,\bullet,\cp \}$ from 
Eq (\ref{eqn:incrementalEnsembleMountainHeight}) we have 
\begin{eqnarray}
\label{eqn:structuralDistanceM}
m(x) &= \left\{ \begin{array}{ll}
1 &\mbox{if $x=\op$}\\
0 &\mbox{if $x=\bullet$} \\
-1 &\mbox{if $x=\cp$}
\end{array} \right.
\end{eqnarray}

Define the distance $d_0(x,y)$ 
between characters $x,y$ in the dot-bracket representation of a secondary
structure by
\begin{eqnarray}
\label{eqn:structuralDistanceD0}
d_0(x,y) = |m(x)-m(y)| &= \left\{ \begin{array}{ll}
0 &\mbox{if $x=y$}\\
1 &\mbox{if [$x=\bullet,y\in \{ \op,\cp \}$] or 
  [$x\in \{ \op,\cp \},y=\bullet$]}\\
2 &\mbox{if [$x=\op,y=\cp$] or [$x=\cp,y=\op$]}
\end{array} \right.
\end{eqnarray}
Let  
$A = \left( \begin{array}{l}
{s_1^* \cdots s_N^*}\\
{t_1^* \cdots t_N^*} 
\end{array} \right)$
denote an alignment between two arbitrary secondary structures $s,t$ of 
(possibly different) lengths $n,m$, where $s_i^*,t_i^* \in
\{ \op,\bullet,\cp,- \}$ and $-$ denotes the gap symbol.
We define the {\em structural alignment distance} for $A$ by summing
$d_0(s_i^*,t_i^*)$ over those positions $i$ where neither character
$s_i^*,t_i^*$ is a gap symbol, then adding $w(k)$ for all size $k$ gaps
in $A$.  Using previous definitions of incremental ensemble expected
mountain height from
Eq (\ref{eqn:incrementalEnsembleMountainHeight}), we can 
generalize structural alignment distance from the simple case of
comparing two dot-bracket representations of secondary structures 
to the more representative case of
comparing the low-energy Boltzmann ensemble of secondary structures for RNA
sequence ${\bf a}$ to that of RNA sequence ${\bf b}$.
Given sequences ${\bf a}= a_1,\ldots,a_n$ and ${\bf b}= b_1,\ldots,b_m$,
let 
$A = \left( \begin{array}{l}
{m_{\bf a}(1)^* \cdots m_{\bf a}(N)^*}\\
{m_{\bf b}(1)^* \cdots m_{\bf b}(N)^*}
\end{array} \right)$
denote an alignment between the incremental ensemble expected mountain height
$m_{\bf a}(1) \cdots m_{\bf a}(n)$ of ${\bf a}$ and and the ensemble 
incremental 
expected mountain height $m_{\bf b}(1) \cdots m_{\bf b}(m)$ of ${\bf b}$.
Generalize structural distance $d_0$ defined in Eq
(\ref{eqn:structuralDistanceD0}) to $d_1$ defined by
$d_1(a_i,b_j) = |m_a(i)-m_b(j)|$, where $m_a(i)$ and $m_b(j)$ are real numbers in the interval $[-1,1]$,
and define {\em ensemble structural alignment distance} for $A$ by summing
$d_1(a_i,b_j)$ over all positions $i,j$ for which neither
character is a gap symbol, then adding positive weight $w(k)$ for all size $k$ 
gaps. By Eq (\ref{eqn:distanceFromSimilarity}) and Eq
(\ref{eqn:distanceGapFromSimilarityGap}), it follows that an
equivalent {\em ensemble structural 
similarity} measure between two positions $a_i,b_j$, denoted $STRSIM(a_i,b_j)$, is obtained by multiplying $d_1$ and
$w(k)$ by $-1$: 
\begin{eqnarray}
STRSIM(a_i,b_j)=-|m_a(i)-m_b(j)|
\end{eqnarray}
This equation will be used later, since
our algorithm {\tt RNAmountAlign} combines both sequence and
ensemble structural similarity.  Indeed, $-|m_a(i)-m_b(j)|\in[-2,0]$ with maximum value of 0 while RIBOSUM85-60, shown in Table \ref{table:RIBOSUM}, has similarity values in the interval $[-1.86,2.22]$. In order to combine sequence with structural similarity, both ranges should be rendered comparable as shown in the next section.

\subsubsection{Pairwise alignment}

\begin{figure*}
\centering
\includegraphics[width=0.8\textwidth]{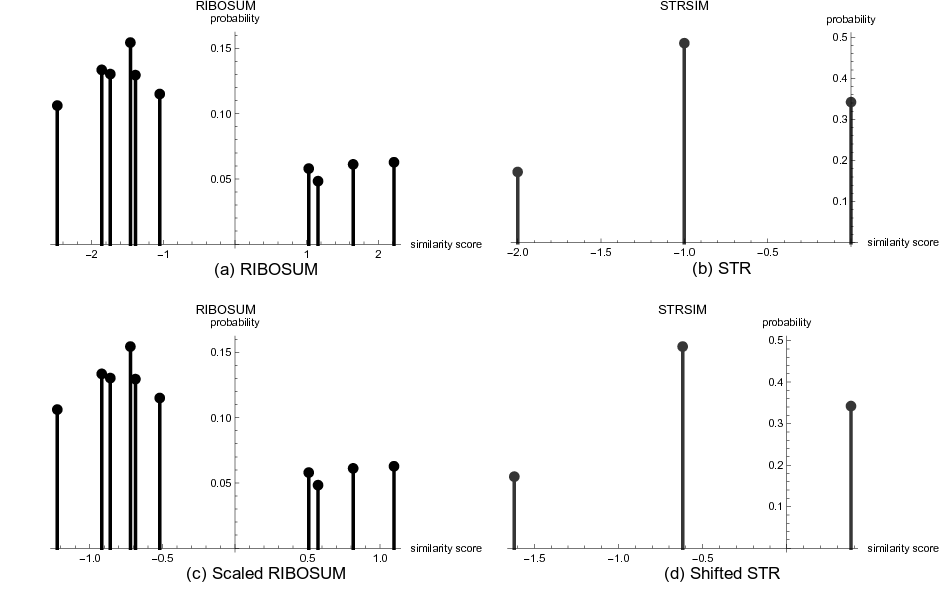}

\caption{For 72 nt tRNA query sequence AL671879.2, nucleotide 
frequencies are approximately
$p_A = 0.167$, $p_C = 0.278$, $p_G = 0.333$, $p_U = 0.222$, and for 
69 nt tRNA target sequence D16498.1, nucleotide
frequencies are approximately
$p_A = 0.377$, $p_C = 0.174$, $p_G = 0.174$, $p_U = 0.275$.
From the base pairing probabilities computed by {\tt RNAfold -p}, we have
query frequencies
$p_{\op} = 0.3035$, $p_{\bullet} = 0.3930$, $p_{\cp} = 0.3035$
and target frequencies
$p_{\op} = 0.2835$, $p_{\bullet} = 0.433$, $p_{\cp} = 0.2835$, so
by Eqs
(\ref{eqn:muSeq},\ref{eqn:sigmaSeq},\ref{eqn:muStr},\ref{eqn:sigmaStr}),
we have
$\mu_{\mbox{\tiny seq}} = -0.9098$, $\sigma_{\mbox{\tiny seq}} = 1.4117$
and
$\mu_{\mbox{\tiny str}} = -0.8301$, $\sigma_{\mbox{\tiny str}} = 0.6968$.
By Eqs (\ref{eqn:alpha0}) and (\ref{eqn:alpha1}), we determine that
RIBOSUM scaling factor 
$\alpha_{\mbox{\tiny seq}} =  0.4936$ and
$\alpha_{\mbox{\tiny str}} = 0.3810$ (values shown only to 4-decimal places).
Panels (a) resp. (b) show the distribution of RIBOSUM resp. STRSIM 
values for the
nucleotide and base pairing probabilities determined from query and target,
while
panels (c) resp. (d) show the distribution of 
$\alpha_{\mbox{\tiny seq}}$-scaled RIBOSUM values resp.
$\alpha_{\mbox{\tiny str}}$-shifted STRSIM values. It follows that
distributions in panels (c) and (d) have the same (negative)
mean and standard deviation.
}
\label{fig:RIBOSUMandSTRSIM}
\end{figure*}

In order to combine sequence and ensemble structural similarity, we
determine a multiplicative scaling factor $\alpha_{\mbox{\tiny seq}}$ 
and an additive shift factor $\alpha_{\mbox{\tiny str}}$ such that the
mean and standard deviation for the distribution of sequence similarity
values from a RIBOSUM matrix \cite{eddy:RSEARCH} (after being multiplied
by $\alpha_{\mbox{\tiny seq}}$) are equal to the
mean and standard deviation for the distribution of structural
similarity values from STRSIM (after additive shift of
$\alpha_{\mbox{\tiny str}}$). The RIBOSUM85-60 nucleotide similarity matrix used in this paper 
is given in Table~\ref{table:RIBOSUM}, and the distributions for
RIBOSUM and STRSIM values are shown in Figure~\ref{fig:RIBOSUMandSTRSIM}
for the 72 nt transfer RNA AL671879.2.
Given query [resp. target] 
nucleotide frequencies $p_A,p_C,p_G,p_U$ [$p'_A,p'_C,p'_G,p'_U$]
that sum to $1$,
the mean $\mu_{\mbox{\tiny seq}}$ and standard deviation 
$\sigma_{\mbox{\tiny seq}}$ of RIBOSUM nucleotide similarities can be 
computed by
\begin{eqnarray}
\label{eqn:muSeq}
\mu_{\mbox{\tiny seq}}  &= \sum\limits_{x,y \in \{A,C,G,U\}} p_x p'_y \cdot
RIBOSUM(x,y)\\
\label{eqn:sigmaSeq}
\sigma_{\mbox{\tiny seq}}  &= \sqrt{\sum\limits_{x,y \in \{A,C,G,U\}} p_x p'_y 
\cdot RIBOSUM(x,y)^2 - \mu_{\mbox{\tiny seq}}^2}
\end{eqnarray}
Setting $s_0(x,y) = -d_0(x,y)$, where $d_0(x,y)$ is
defined in Eq (\ref{eqn:structuralDistanceD0}),
for given query [resp. target] base pairing probabilities 
$p_{\op},p_{\bullet},p_{\cp}$ [resp.  $p'_{\op},p'_{\bullet},p'_{\cp}$]
of
dot-bracket characters, it follows that
the mean $\mu_{\mbox{\tiny str}}$ and standard deviation 
$\sigma_{\mbox{\tiny str}}$ of structural similarities can be computed by
\begin{eqnarray}
\label{eqn:muStr}
\mu_{\mbox{\tiny str}}  &= \sum\limits_{x,y \in \{\op,\bullet,\cp\}} 
p_x p'_y \cdot s_0(x,y)\\
\label{eqn:sigmaStr}
\sigma_{\mbox{\tiny str}}  &= \sqrt{\sum\limits_{x,y \in \{\op,\bullet,\cp\}}
p_x p'_y \cdot s_0(x,y)^2 - \mu_{\mbox{\tiny str}}^2}
\end{eqnarray}
Now we compute a multiplicative
factor $\alpha_{\mbox{\tiny seq}}$ and an 
additive shift term $\alpha_{\mbox{\tiny str}}$, both dependent
on frequencies $p_A,p_C,p_G,p_U$ and $p_{\op},p_{\bullet},p_{\cp}$,
such that the mean [resp. standard deviation]
of nucleotide similarity multiplied by $\alpha_{\mbox{\tiny seq}}$ 
is equal to the mean [resp. standard deviation]
of structural similarity after addition of shift term 
$\alpha_{\mbox{\tiny str}}$:

\begin{eqnarray}
\label{eqn:alpha0}
\alpha_{\mbox{\tiny seq}} &=
\sigma_{\mbox{\tiny str}}/ \sigma_{\mbox{\tiny seq}} \\
\label{eqn:alpha1}
\alpha_{\mbox{\tiny str}} &= 
\alpha_{\mbox{\tiny seq}} \cdot
\mu_{\mbox{\tiny seq}} - \mu_{\mbox{\tiny str}} 
\end{eqnarray}


\begin{table}
\centering
\begin{small}
\begin{tabular}{|l|c c c c|}
\hline
& A & C & G & U \\
\hline
A & +2.22 & -1.86 & -1.46 &  -1.39 \\
C & -1.86 & +1.16 & -2.48 & -1.05 \\
G & -1.46 & -2.48 & +1.03 & -1.74 \\
U & -1.39 & -1.05 & -1.74 & +1.65 \\
\hline
\end{tabular}
\end{small}
\caption{RIBOSUM85-60 similarity matrix for RNA nucleotides from \cite{eddy:RSEARCH}.
}
\label{table:RIBOSUM}
\end{table}

Given the query RNA 
${\bf a} = a_1,\ldots,a_n$ and target RNA ${\bf b} = b_1,\ldots,b_m$ with
incremental ensemble expected mountain heights
$m_{\bf a}(1) \cdots m_{\bf a}(m)$ of ${\bf a}$,
$m_{\bf b}(1) \cdots m_{\bf b}(m)$ of ${\bf b}$, and
user-defined weight $0 \leq \gamma \leq 1$,
our final similarity measure is defined by
\begin{eqnarray}
\label{eqn:similarityMeasure}
\mbox{sim}_{\gamma}(a_i,b_j) &=
(1-\gamma) \cdot \alpha_{\mbox{\tiny seq}} \cdot RIBOSUM(a_i,b_j) \\
\nonumber
&+ \gamma \cdot \left(
\alpha_{\mbox{\tiny str}} + STRSIM(a_i,b_j) \right)
\end{eqnarray}
where
$\alpha_{\mbox{\tiny seq}}, \alpha_{\mbox{\tiny str}}$ are computed by
Eqs (\ref{eqn:alpha0},\ref{eqn:alpha1}) depending on 
probabilities $p_A,p_C,p_G,p_U$ [resp. $p'_A,p'_C,p'_G,p'_U$] and
$p_{\op},p_{\bullet},p_{\cp}$ [resp. $p'_{\op},p'_{\bullet},p'_{\cp}$]
of the query [resp. target].
All benchmarking computations were carried out using
$\gamma=1/2$, although it is possible to use position-specific weight
$\gamma_{i,j}$ defined as the average probability that
$i$ is paired in ${\bf a}$ and $j$ is paired in ${\bf b}$. 

Our structural similarity measure is closely related to that of
{\tt STRAL}, which we discovered only after completing a preliminary
version of this paper. Let 
$pl^a_i= \sum_{j<i}{p^a_{j,i}}$ and $pr^a_i= \sum_{j>i}{p^a_{i,j}}$ 
be the probability that position $i$ of sequence {\bf a} is paired to 
a position on the left or right, respectively. The similarity measure
used in {\tt STRAL} is defined by
\begin{align}
\label{eqn:similaritySTRAL}
\mbox{sim}^{STRAL}_{\gamma}(a_i,b_j) &= \gamma \cdot \big( \sqrt{pl^a_i \cdot pl^b_j} +  \sqrt{pr^a_i \cdot pr^b_j}\big) \nonumber \\
&+ \sqrt{(1 -pr^a_i - pl^a_i) \cdot (1 -pr^a_i - pl^a_i)} \cdot RIBOSUM(a_i,b_j)
\end{align}

From Eq (\ref{eqn:similarityMeasure}) and  Eq
(\ref{eqn:incrementalEnsembleMountainHeight}) our measure can be defined as
\begin{eqnarray}
\label{eqn:similarityMeasureSTRAL}
\mbox{sim}_{\gamma}(a_i,b_j) &=
\gamma \cdot \left(
\alpha_{\mbox{\tiny str}} - |(pr^a_i - pl^a_i) -  (pr^b_j - pr^b_j) | \right)
\nonumber \\
&+ (1-\gamma) \cdot \alpha_{\mbox{\tiny seq}} \cdot RIBOSUM(a_i,b_j)
\end{eqnarray}
Though {\tt RNAmountAlign} was developed independently much later than 
{\tt STRAL}, our software offers functionalities unavailable in {\tt STRAL},
which latter appears to be no longer maintained.\footnote{Since we were 
unable to compile {\tt STRAL}, our benchmarking results for {\tt STRAL} use
an adaptation of our code to support Eq (\ref{eqn:similaritySTRAL}).
There are nevertheless some differences in how progressive
alignment is implemented in {\tt STRAL} that could affect run time.}
For instance, {\tt RNAmountAlign} supports local and semiglobal alignment,
and reports $p$-values and E-values; these features are not available in
{\tt STRAL}.

To illustrate the method, suppose that the query [resp. target] sequence 
is the 72 nt tRNA AL671879.2 [resp. 69 nt tRNA D16498.1]. Then
nucleotide query [resp. target] probabilities are (approximately)
$p_A = 0.167$, $p_C = 0.278$, $p_G = 0.333$, $p_U = 0.222$,
[resp.  $p'_A = 0.377$, $p'_C = 0.174$, $p'_G = 0.174$, $p'_U = 0.275$].
From the base pairing probabilities returned by {\tt RNAfold -p} 
\cite{Lorenz.amb11}, we determine that
$p_{\op} = 0.3035$, $p_{\bullet} = 0.3930$, $p_{\cp} = 0.3035$
[resp.  $p'_{\op} = 0.2835$, $p'_{\bullet} = 0.433$, $p'_{\cp} = 0.2835$].
Using these probabilities in Eqs
(\ref{eqn:muSeq}--\ref{eqn:sigmaStr}),
we determine that
$\mu_{\mbox{\tiny seq}} = -0.9098$, $\sigma_{\mbox{\tiny seq}} = 1.4117$,
and 
$\mu_{\mbox{\tiny str}} = -0.8301$, $\sigma_{\mbox{\tiny str}} = 0.6968$.
By Eq (\ref{eqn:alpha0}) and Eq (\ref{eqn:alpha1}), we determine that
RIBOSUM scaling factor $\alpha_{\mbox{\tiny seq}} =  0.4936$ and
$\alpha_{\mbox{\tiny str}} = 0.3810$. It follows that
the mean and standard deviation of 
$\alpha_{\mbox{\tiny seq}}$-scaled RIBOSUM values are identical with that of
$\alpha_{\mbox{\tiny str}}$-shifted STRSIM values, hence can be combined
in Eq (\ref{eqn:similarityMeasure}).  Since 
sequence identity of the {\tt BRAliBase 2.1} alignment of these tRNAs is
only 28\%, we set structural similarity weight $\gamma=1$ in 
Eq (\ref{eqn:similarityMeasure}), and obtained a (perfect)
global alignment computed by {\tt RNAmountAlign}. 
Figure~\ref{fig:RIBOSUMandSTRSIM} depicts the distribution of
RIBOSUM85-60 [resp. STRSIM] values in this case, both
before and after application of scaling factor
$\alpha_{\mbox{\tiny seq}}$ [resp. shift
$\alpha_{\mbox{\tiny str}}$] -- recall that
$\alpha_{\mbox{\tiny seq}}$ and
$\alpha_{\mbox{\tiny str}}$] depend on
$p_A,p_C,p_G,p_U,p_{\op},p_{\bullet},p_{\cp}$ of
tRNA AL671879.2 and
$p'_A,p'_C,p'_G,p'_U,p'_{\op},p'_{\bullet},p'_{\cp}$ of
tRNA D16498.1.

\subsubsection{Statistics for pairwise alignment}

\textbf{Karlin-Altschul statistics for local pairwise alignment.} For a finite alphabet $A$ and similarity measure $s$, suppose that the
expected similarity $\sum\limits_{x,y \in A} p_x p_y \cdot s(x,y)$ is
negative and that $s(x,y)$ is positive for at least one choice of $x,y$.
In the case of {\tt BLAST}, amino acid and nucleotide similarity
scores are integers, for which the Karlin-Altschul algorithm was developed \cite{Karlin.pnas90}. In contrast, {\tt RNAmountAlign} similarity scores
scores are not integers (or more generally values in a lattice), 
because Eq (\ref{eqn:similarityMeasure}) combines
real-valued $\alpha_{\mbox{\tiny seq}}$-scaled
RIBOSUM nucleotide similarities with real-valued 
$\alpha_{\mbox{\tiny str}}$-shifted
STRSIM structural similarities, which depend on
query [resp. target] probabilities 
$p_A,p_C,p_G,p_U,p_{\op},p_{\bullet},p_{\cp}$ 
[resp. $p'_A,p'_C,p'_G,p'_U,p'_{\op},p'_{\bullet},p'_{\cp}$].
For that reason, we use the following reformulation of a
result by Karlin, Dembo and Kawabata \cite{karlinDemboKawabata}, 
the similarity score $s(x,y)$ for RNA nucleotides $x,y$ is
defined by Eq (\ref{eqn:similarityMeasure}). 

\noindent
\begin{theorem}[Theorem 1 of \cite{karlinDemboKawabata}]\hfill\break
Given similarity measure $s$ between nucleotides in alphabet
$A=\{A,C,G,U\}$, let $\lambda^*$ be the unique positive root of 
$E[e^{s(x,y)}] = \sum\limits_{x,y \in A} p_x p'_y \cdot e^{\lambda s(x,y)}$,
and let random variable $S_k$ denote the score of a length $k$ gapless
alignment.  For large $z$,
\begin{align*}
P\left( M > \frac{\ln n m}{\lambda^*} + z
\right) &\leq \exp(-K^* e^{-\lambda^*z})
\end{align*}
where $M$ denotes high maximal segment scores for local alignment
of random RNA sequences $a_1,\ldots,a_n$ and $b_1,\ldots,b_m$, and where
\begin{align*}
K^* &= \frac{\exp\left( -2 \sum_{k=1}^{\infty} \frac{1}{k} \cdot
(E[e^{\lambda^* S_k; S_k<0}] + P(S_k\geq 0)
\right)}
{\lambda^* E[Xe^{\lambda^* X}]}
\end{align*}
\end{theorem}

\textbf{Fitting data to probability distributions.} Data were fit to the normal distribution (ND) by the method of moments
(i.e. mean and standard deviation were taken from data analysis).
Data were fit to the extreme value distribution (EVD) 
\begin{eqnarray}
P(x<s) &= 1-\exp( -Ke^{\lambda s} )
\end{eqnarray}
by an in-house
implementation of maximum likelihood to determine $\lambda,K$, 
as described in supplementary information to \cite{eddy:RSEARCH}.
Data were fit to the gamma distribution by using the function
{\tt fitdistr(x,'gamma')} from the package {\tt MASS} in the
{\tt R} programming language, which determines
rate and shape parameters for the density function
\begin{eqnarray}
f(x,\alpha,\lambda) &= \frac{\lambda^{\alpha} x^{\alpha-1} e^{-\lambda x}}
{\Gamma(\alpha)}
\end{eqnarray}
 with 
where $\alpha$ is the shape parameter, the rate is $1/\lambda$, where 
$\lambda$ is known as the scale parameter.  

\subsubsection{Multiple alignment}
Suppose $p_A,p_C,p_G,p_U$ are the nucleotide probabilities obtained after the concatenation of all sequences. Let $p_{\op},p_{\bullet},p_{\cp}$ be computed by individually folding each sequence and taking the arithmetic average of probabilities of $\op$ , $\bullet$ and $\cp$ over all sequences. The mean and standard deviation of sequence and structure similarity are computed similar to 
Eqs (\ref{eqn:muSeq}-\ref{eqn:sigmaStr}).

\begin{eqnarray}
\label{eqn:muSeqMulti}
\mu_{\mbox{\tiny seq}}  &= \sum\limits_{x,y \in \{A,C,G,U\}} p_x p_y \cdot
RIBOSUM(x,y)\\
\label{eqn:sigmaSeqMulti}
\sigma_{\mbox{\tiny seq}}  &= \sqrt{\sum\limits_{x,y \in \{A,C,G,U\}} p_x p_y 
\cdot RIBOSUM(x,y)^2 - \mu_{\mbox{\tiny seq}}^2}
\end{eqnarray}

\begin{eqnarray}
\label{eqn:muStrMulti}
\mu_{\mbox{\tiny str}}  &= \sum\limits_{x,y \in \{\op,\bullet,\cp\}} 
p_x p_y \cdot s_0(x,y)\\
\label{eqn:sigmaStrMulti}
\sigma_{\mbox{\tiny str}}  &= \sqrt{\sum\limits_{x,y \in \{\op,\bullet,\cp\}}
p_x p_y \cdot s_0(x,y)^2 - \mu_{\mbox{\tiny str}}^2}
\end{eqnarray}
Sequence multiplicative scaling factor $\alpha_{\mbox{\tiny seq}}$ 
and the structure additive shift factor $\alpha_{\mbox{\tiny str}}$ 
are computed from these values using Eqs ~(\ref{eqn:alpha0},\ref{eqn:alpha1}).

{\tt RNAmountAlign} implements progressive multiple alignment using {\tt UPGMA} to construct the guide tree. In {\tt UPGMA}, one first defines a similarity matrix $S$, where $S[i,j]$ is equal to (maximum) pairwise sequence similarity of sequences $i$ and $j$. A rooted tree is then constructed by progressively creating a parent node of the two closest siblings. Parent nodes are profiles (PSSMs) that represent alignments of two or more sequences, hence can be treated as pseudo-sequences in a straightforward adaptation of pairwise alignment to the alignment of profiles.
Let's consider an alignment of $N$ sequences $A = \left( \begin{array}{c}
{a_{11}^*  \cdots a_{1M}^*}\\
 \cdots \\
{a_{N1}^*  \cdots a_{NM}^*} 
\end{array} \right)$ composed of $M$ columns. Let $A_{i} = \{\aseq^*_{1i},\aseq^*_{2i},\ldots,\aseq^*_{Ni}\}$ denote column $i$ of the alignment (for $1\leq i \leq M$). Suppose $p(i,x)$, for $x\in\{A,C,G,U,-\}$, indicates the probability of occurrence of a nucleotide or gap at column $i$ of alignment $A$. Then sequence similarity SEQSIM between two columns is defined by

\begin{eqnarray}
\label{eqn:seqsimMulti}
SEQSIM(A_{i},A_{j}) = \sum_{x \in \{A,C,G,U,-\}} \sum_{y \in \{A,C,G,U,-\}} p(i,x)\cdot p(j,y) \cdot R(x,y)
\end{eqnarray}
where
\begin{eqnarray}
\label{eqn:riboMulti}
R(x,y) &= \left\{ \begin{array}{ll}
0 &\mbox{if $x=-$ or $y=-$}\\
RIBOSUM(x,y) &\mbox{otherwise}\\
\end{array} \right.
\end{eqnarray}

The structural measure for a profile is computed from the incremental ensemble heights averaged over each column. Let $m_A(i)$ denote the arithmetic average of incremental ensemble mountain height at column $A_i$
\begin{eqnarray}
m_A(i) =\frac{\sum_{1\leq j \leq N} m_{\aseq^*_{j}}(i)}{N}
\label{eqn:maMulti}
\end{eqnarray}   
where $m_{\aseq^*_{j}}(i)$ is the incremental ensemble mountain height at position $i$ of sequence $\aseq^*_{j}$ obtained from 
Eq~(\ref{eqn:incrementalEnsembleMountainHeight}).
Here, let $m_{\aseq^*_{j}}(i)=0$ if $\aseq^*_{ji}$ is a gap. Structural similarity between two columns is defined by
\begin{eqnarray}
STRSIM(A_i,A_j)=-|m_A(i) - m_A(j)|
\label{eqn:strSimMulti}
\end{eqnarray}
Finally, the combined sequence/structure similarity is computed from
\begin{eqnarray}
\mbox{sim}_{\gamma}(A_i,A_j) &=
(1-\gamma) \cdot \alpha_{\mbox{\tiny seq}} \cdot SEQSIM(A_i,A_j) \\
\nonumber
&+ \gamma \cdot \left(
\alpha_{\mbox{\tiny str}} + STRSIM(A_i,A_j) \right)
\label{eqn:simMulti}
\end{eqnarray}

\subsection{Benchmarking}

\subsubsection{Accuracy measures}

Sensitivity, positive predictive value, and F1-measure for pairwise 
alignments were computed as follows.  Let
$A = \left( \begin{array}{l}
{a_1^* \cdots a_n^*}\\
{b_1^* \cdots b_n^*}
\end{array} \right)$
denotes an alignment, where $a_i,b_i \in 
\{ A,C,G,U,\mbox{\textemdash}\}$, and the aligned sequences 
include may contain gap symbols $\mbox{\textemdash}$ provided that
it is not the case that both $a^*_i$ and $b^*_i$ are gaps. The number TP
of true positives [resp. FP of false positives] is the number of
alignment pairs $(a^*_i,b^*_i)$ in the predicted alignment that belong to
[resp. do not belong to] the reference alignment. The sensitivity ($Sen$)
[resp. positive predictive value ($PPV$)] of a predicted alignment is
TP divided by reference alignment length
[resp. TP divided by predicted alignment length].
The $F1$-score is the harmonic mean of
sensitivity and PPV, so $F1 = \frac{2}{1/Sen + 1/PPV}$. For the computation of $Sen$ ,$PPV$, and $F1$, pairs of the form $(X,\mbox{\textemdash})$ and $(\mbox{\textemdash},X)$ are also counted. 
In the case of local alignment, since the size of the reference alignment 
is unknown, only the predicted alignment length and PPV are reported. 
To compute the accuracy of multiple alignment, we used sum-of-pair-scores 
(SPS) \cite{Thompson.nar99}, defined as follows. Suppose that $A$ denotes
a multiple alignment of the form
$A = \left( \begin{array}{c}
{a_{11}^*  \cdots a_{1M}^*}\\
 \cdots \\
{a_{N1}^*  \cdots a_{NM}^*}
\end{array} \right)$.
For $1 \leq i,j \leq M$, $1 \leq k \leq N$ define $p_{ijk}=1$ 
if $a^*_{ik}$ is aligned with $a^*_{jk}$ in both the reference and 
predicted alignments, and $p_{ijk}=0$ otherwise. Sum-of-pairs score
SPS is then the sum, taken over all $i,j,k$, of the $p_{ijk}$.
Though SPS can be considered as the average sensitivity, taken over 
all sequence pairs in the alignment, this is not technically the case,
since our definition of sensitivity also counts pairs of the form
$(X,\mbox{\textemdash})$ and
$(\mbox{\textemdash},X)$ from the reference alignment. 

To measure the conservation of secondary structures in alignments, 
structural conservation index (SCI) was computed using {\tt RNAalifold} \cite{Bernhart.bb08}. {\tt RNAalifold} computes SCI as the ratio of the 
free energy of the alignment, computed by {\tt RNAalifold}, with the
average minimum free energy of individual structures in the alignment.
SCI values close to $1$  [resp. $0$] indicate high [resp. low]
structural conservation. All computations made with Vienna RNA Package
used version 2.1.7 \cite{Lorenz.amb11} using default Turner 2004 energy 
parameters \cite{Turner.nar10}). 

\subsubsection{Dataset for global and local alignment comparison} 

For  {\em pairwise global} alignment benchmarking in
Table~\ref{table:F1measures} and 
Figures \ref{fig:FSCI}, \ref{fig:runTime}, \ref{fig:sensglobalAlignmentRfam} and \ref{fig:PPVglobalAlignmentRfam}
all 8976 pairwise alignments in k2 from {\tt BRAliBase 2.1} database \cite{Gardner.nar05} were used. For  {\em multiple global} alignment 
benchmarking in Fig ~\ref{fig:multiFSCI}, k5 {\tt BRAliBase 3} 
was used \cite{Freyhult.gr07}. This dataset includes 583 reference 
alignments, each composed of 5 sequences. 
For \textit{pairwise local alignment} benchmarking, 
75 pairwise alignments having sequence identity $\leq 70\%$ were 
randomly selected
from each of 20 well-known families from the
Rfam 12.0 database \cite{Nawrocki.nar15}, many of which were
considered in a previous study \cite{Clote.r05}, yielding a total of
1500 alignments. Following \cite{Tabei.b09},
these alignments were trimmed on the left and right, so that
both first and last aligned pairs of the alignment do not contain a gap
symbol. For sequences 
${\bf a} = a_1,\ldots,a_n$ [resp. ${\bf b}=b_1,\ldots,b_m$] from each alignment,
random sequences ${\bf a}'$ [resp. ${\bf b}'$] were generated with the same 
nucleotide frequencies, then a random position was chosen in
${\bf a}'$ [resp. ${\bf b}'$] in which to insert ${\bf a}$ [resp. ${\bf b}$], 
thus resulting in a pair of sequences of lengths $4n$ and $4m$.  Finally,
since sequence identity was at most 70\%, the
RIBOSUM70-25 similarity matrix was used in {\tt RNAmountAlign}.
Preparation of the benchmarking dataset for local alignment was analogous to
the method used in {\em multiple} local alignment of \cite{Tabei.b09}. We used {\tt LocARNA} (version 1.8.7), {\tt FOLDALIGN} (version 2.5), 
{\tt LARA} (version 1.3.2)
{\tt DYNALIGN} (from version 5.7 of {\tt RNAstructure}), and {\tt STRAL} (in-house implementation due to unavailability) for benchmarking.

\subsubsection{Dataset for correlation of $p$-values for different
distribution fits}

A pool of 2220 sequences from the Rfam 12.0 database \cite{Nawrocki.nar15}
was created as follows. One sequence was selected from each
Rfam family having average sequence length at most $200$ nt, 
with the property that the base pair distance
between its minimum free energy (MFE) structure and the Rfam consensus 
structure was a minimum.  Subsequently,
for each of 500 randomly selected {\em query} sequences from the pool 
of 2220 sequences, 1000 random {\em target} sequences of length 400 nt
were generated 
to have the same expected nucleotide frequency as that of the query.  
For each query and random target,
five semiglobal (query search) alignments were created using gap 
initiation costs
of $g_i \in \{ -1,-2,-3,-4,-5 \}$ with gap extension cost $g_e$ equal to
one-third the gap initiation cost.  For each 
alignment score $x$ for query and random target, the $p$-value was computed as
$1-CDF(x)$ for ND, EVD and GD, where $CDF(x)$ is the cumulative density function
evaluated at $x$. Additionally, a heuristic $p$-value was determined by
calculating the proportion of alignment scores for given query that exceed
$x$. 

\section{Results}
We benchmarked {\tt RNAmountAlign}'s performance for pairwise and multiple alignments on {\tt BraliBase} k2 and k5 datasets, respectively.
\subsection{Pairwise alignment}
\begin{figure*}
\centering
\includegraphics[width=0.8\textwidth]{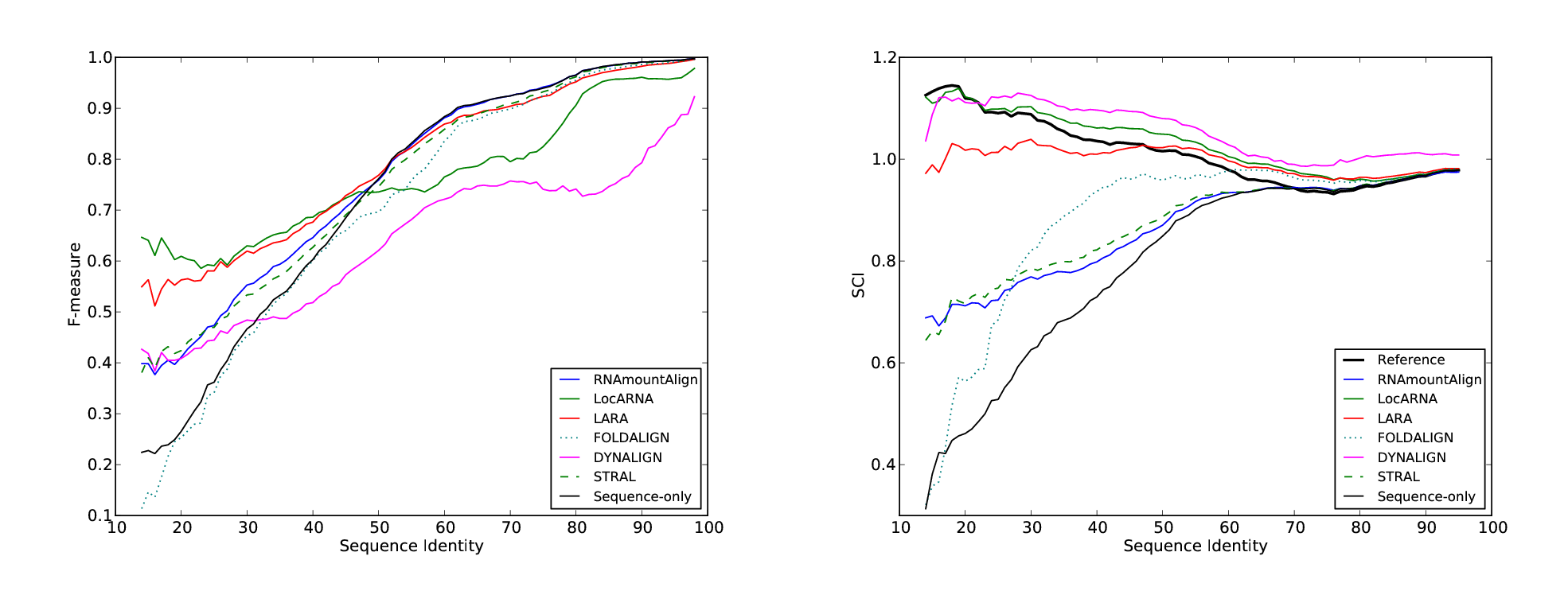}

\caption{F1-measure (Left) and structural conservation index (SCI) (Right) for \textit{pairwise global  alignments} using 
{\tt RNAmountAlign}, {\tt LocARNA}, {\tt LARA}, {\tt FOLDALIGN},
{\tt DYNALIGN}, {\tt STRAL} and sequence-only($\gamma=0$). 
F1-measure and SCI are shown as a function of 
alignment sequence identity for pairwise alignments in the
{\tt BRAliBase 2.1} database used for benchmarking.
}
\label{fig:FSCI}
\end{figure*}

Figures  \ref{fig:FSCI}, \ref{fig:sensglobalAlignmentRfam} and \ref{fig:PPVglobalAlignmentRfam} depict
running averages of {\em pairwise global alignment}
F1-measure, sensitivity, and positive predictive value (PPV)
for the software described in this paper, as well
as for {\tt LocARNA}, {\tt FOLDALIGN}, 
{\tt LARA},
{\tt DYNALIGN}, and {\tt STRAL}.
For pairwise benchmarking, reference alignments of size 2, a.k.a. K2, were
taken from the {\tt BRAliBase 2.1} database \cite{Gardner.nar05}. 
{\tt BRAliBase 2.1 K2} data are based on seed alignments of the Rfam 7.0 
database, and consist of 8976 alignments of RNA sequences from 36 Rfam families.

\begin{figure*}
\centering
\includegraphics[width=0.8\textwidth]{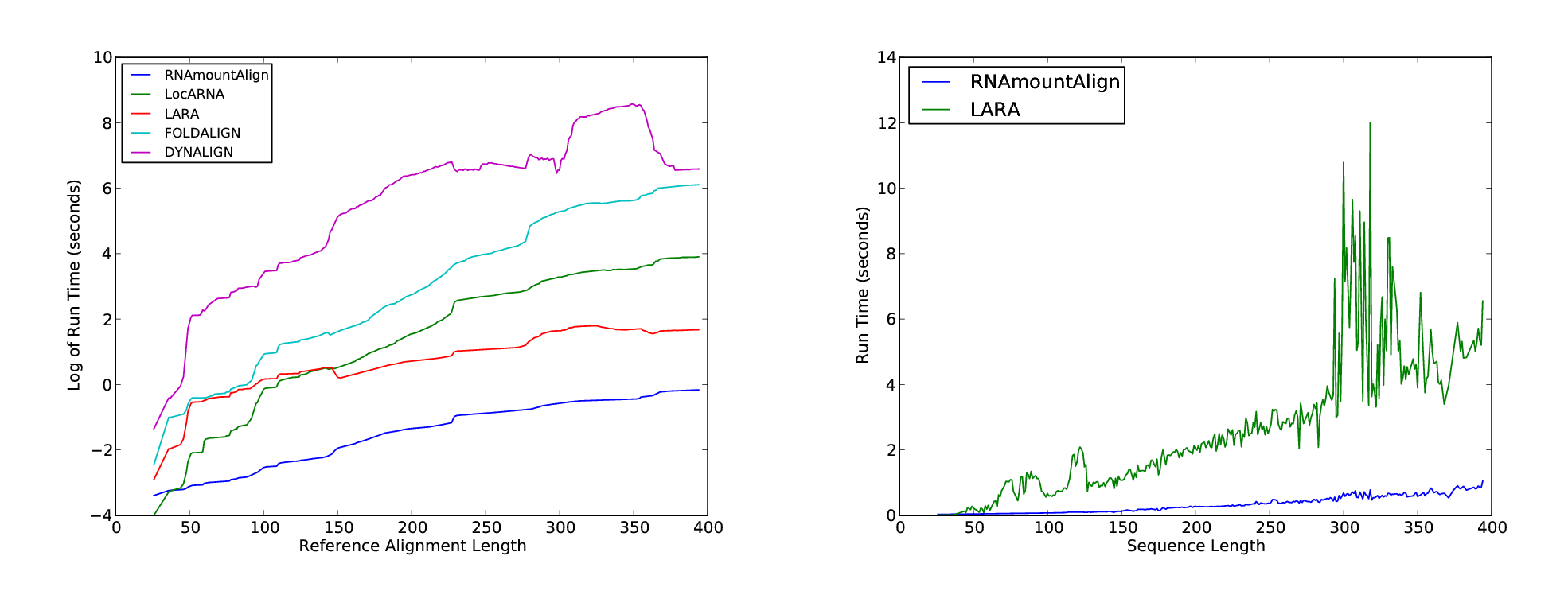}

\caption{Run time of \textit{pairwise global alignment} for
{\tt RNAmountAlign}, {\tt LocARNA}, {\tt LARA}, {\tt FOLDALIGN}, and
{\tt DYNALIGN}. 
(Left) Log run time is shown as a function of 
seed length for pairwise alignments in the
{\tt BRAliBase 2.1} database used for benchmarking. Window size of 51 is used for the computation of moving average.
(Right) Actual run time for {\tt RNAmountAlign} and {\tt LARA} on 
the same data. Unlike
the left panel the actual run time is
shown, rather than log run time, without any moving average taken.
}
\label{fig:runTime}
\end{figure*}

Running averages of sensitivity, positive predictive value, and F1-measure,
averaging over windows of size 11 nt (interval $[k-5,k+5]$), 
were computed as a function of sequence
identity, where it should be noted that the number of pairwise alignments
for different values of sequence identity can vary for the {\tt BRAliBase 2.1} 
data (e.g. there are only 35 pairwise alignments having sequence identity 
$<20\%$). Default parameters were used for all other software.  For our software {\tt RNAmountAlign}, gap initiation cost was
-3, gap extension -1, and sequence/structure weighting parameter $\gamma$
was $0.5$ (value obtained by optimizing on a small set of 300 random alignments from Rfam 12.0, not considered in training or testing set). The sequence-only alignment is computed from {\tt RNAmountAlign} with the same gap penalties, but for $\gamma=0$.
While its accuracy is high,
{\tt RNAmountAlign} is faster by an order of magnitude
than {\tt LocARNA}, {\tt LARA}, 
{\tt FOLDALIGN}, and {\tt DYNALIGN} -- indeed,
algorithmic time complexity of our method is $O(n^3)$ compared with 
$O(n^4)$ for these methods. Since {\tt STRAL} could not be compiled on any of our systems, we implemented its algorithm by modifying {\tt RNAmountAlign} and obtained results for {\tt STRAL}'s default parameter settings. Therefore, the run time of {\tt STRAL} is identical to {\tt RNAmountAlign} but we achieve slightly higher F1-measure, sensitivity and PPV. Moreover, {\tt RNAmountAlign} supports semiglobal and local alignments as well as reporting $p$-values. The right  panel of Fig ~\ref{fig:runTime} depicts actual run times of
the fastest software, {\tt RNAmountAlign},  with the next fastest
software, {\tt LARA}. Unlike the graph in the
left panel, actual run times are shown, graphed as a function of sequence length,
rather than logarithms of moving averages. 

In addition, Table~\ref{table:F1measures} displays average
pairwise global alignment F1 scores for 
{\tt RNAmountAlign}, {\tt LocARNA}, {\tt LARA}, 
{\tt FOLDALIGN}, {\tt DYNALIGN}, and {\tt STRAL} when benchmarked on 36 families
from the {\tt BRaliBase K2} database comprising altogether 8976 RNA sequences
with average length of $249.33$. Averaging over all sequences, the F1
scores for the programs just mentioned were respectively
$0.8370$, $0.7808$, $0.8406$, $0.7977$, $0.6822$, $0.8247$; i.e. F1 score 
$0.8406$ of {\tt LARA} slightly exceeded the F1 score $0.8370$
of {\tt RNAmountAlign} and $0.8247$ of {\tt STRAL}, while other methods trailed by several percentage
points. Supplementary Information (SI) Tables ~\ref{table:globalSensitivity} and ~\ref{table:globalPPV} display values for
global alignment sensitivity and positive predictive value, benchmarked on the 
same data for the same programs -- these results are similar to the F1-scores 
in Tables \ref{table:softwareOverview} and \ref{table:F1measures}. 

Although there appears to be no universally
accepted criterion for quality of local alignments, 
Table ~\ref{table:localAlignmentBenchmarking} shows pairwise 
local alignment comparisons for the above-mentioned methods supporting 
local alignment: {\tt RNAmountAlign}, {\tt FOLDALIGN}, and {\tt LocARNA}. 
We had intended to include {\tt SCARNA\_LM} \cite{Tabei.b09} in the
benchmarking of multiple local alignment software; however, {\tt SCARNA\_LM}
no longer appears to be maintained, since the web server is no longer 
functional and no response came from our request for the source code.
Since the reference alignments for the local benchmarking dataset are not
known, and sensitivity depends upon the length of the reference alignment, 
we only report local alignment length and positive predictive value.
Abbreviating
{\tt RNAmountAlign} by MA, {\tt FOLDALIGN} by FA, and {\tt LocARNA} by LOC,
Table \ref{table:localAlignmentBenchmarking} shows average run time in 
seconds of MA ($2.30\pm 2.12$), FA ($625.53 \pm 2554.61$), 
LOC ($5317.96 \pm 8585.19$), average alignment length of reference alignments 
($118.67 \pm 47.86$), MA ($50.35 \pm 42.33$), FA ($114.86 \pm 125.33$),
LOC ($556.82 \pm 227.00$), and average PPV scores MA ($0.53 \pm 0.42$), 
FA ($0.64 \pm 0.36$), LOC ($0.03 \pm 0.04$).

\begin{table*}
\centering
\resizebox{\textwidth}{0.4\textheight}{
\begin{tabular}{|l|cccccccc|}
\hline 
Type&	NumAln&	SeqId&	MA(F)&	LocARNA(F)&	LARA(F)&	FA(F)&	DA(F)& STRAL(F)\\
\hline 
5.8S rRNA&	$76$&	$0.72\pm 0.13$&	$0.90\pm 0.09$&	$0.82\pm 0.07$&	$0.87\pm 0.15$&	$0.89\pm 0.11$&	$0.66\pm 0.22$&	$ 0.88 \pm 0.12 $\\
5S rRNA&	$1162$&	$0.60\pm 0.14$&	$0.84\pm 0.16$&	$0.87\pm 0.13$&	$0.85\pm 0.16$&	$0.86\pm 0.14$&	$0.69\pm 0.17$&	$ 0.82 \pm 0.20  $\\
Cobalamin&	$188$&	$0.43\pm 0.10$&	$0.56\pm 0.16$&	$0.38\pm 0.17$&	$0.49\pm 0.20$&	$0.43\pm 0.24$&	$0.36\pm 0.19$&	$ 0.54 \pm 0.17 $\\
Entero 5 CRE&	$48$&	$0.88\pm 0.06$&	$0.98\pm 0.04$&	$0.99\pm 0.04$&	$0.99\pm 0.05$&	$0.99\pm 0.02$&	$0.87\pm 0.13$&	$ 0.97 \pm 0.06 $\\
Entero CRE&	$65$&	$0.80\pm 0.07$&	$1.00\pm 0.00$&	$0.99\pm 0.03$&	$0.96\pm 0.07$&	$0.99\pm 0.04$&	$0.76\pm 0.17$&	$ 1.00 \pm 0.03  $\\
Entero OriR&	$49$&	$0.84\pm 0.06$&	$0.95\pm 0.07$&	$0.92\pm 0.09$&	$0.94\pm 0.08$&	$0.94\pm 0.07$&	$0.84\pm 0.15$&	$ 0.95 \pm	0.07 $\\
gcvT&	$167$&	$0.44\pm 0.13$&	$0.61\pm 0.19$&	$0.61\pm 0.24$&	$0.57\pm 0.25$&	$0.40\pm 0.33$&	$0.44\pm 0.19$&	$  0.62\pm	0.20 $\\
Hammerhead 1&	$53$&	$0.71\pm 0.17$&	$0.89\pm 0.13$&	$0.90\pm 0.11$&	$0.87\pm 0.16$&	$0.83\pm 0.25$&	$0.52\pm 0.27$&	$ 0.88\pm	0.16   $\\
Hammerhead 3&	$126$&	$0.66\pm 0.21$&	$0.86\pm 0.20$&	$0.88\pm 0.21$&	$0.88\pm 0.20$&	$0.80\pm 0.31$&	$0.71\pm 0.31$&	$ 0.90\pm	0.16   $\\
HCV SLIV&	$98$&	$0.85\pm 0.05$&	$0.99\pm 0.03$&	$0.98\pm 0.04$&	$0.98\pm 0.03$&	$0.99\pm 0.03$&	$0.81\pm 0.34$&	$  0.99\pm	0.03  $\\
HCV SLVII&	$51$&	$0.83\pm 0.09$&	$0.97\pm 0.06$&	$0.96\pm 0.06$&	$0.93\pm 0.10$&	$0.95\pm 0.07$&	$0.71\pm 0.22$&	$  0.95\pm	0.07  $\\
HepC CRE&	$45$&	$0.86\pm 0.06$&	$1.00\pm 0.00$&	$1.00\pm 0.00$&	$1.00\pm 0.00$&	$1.00\pm 0.00$&	$0.77\pm 0.29$&	$  1.00\pm	0.00  $\\
Histone3&	$84$&	$0.78\pm 0.09$&	$1.00\pm 0.00$&	$1.00\pm 0.00$&	$1.00\pm 0.00$&	$1.00\pm 0.00$&	$1.00\pm 0.00$&	$  1.00\pm	0.00  $\\
HIV FE&	$733$&	$0.87\pm 0.04$&	$1.00\pm 0.02$&	$1.00\pm 0.02$&	$0.98\pm 0.05$&	$0.99\pm 0.05$&	$0.64\pm 0.29$&	$  1.00\pm	0.02  $\\
HIV GSL3&	$786$&	$0.86\pm 0.04$&	$0.99\pm 0.02$&	$0.99\pm 0.02$&	$0.98\pm 0.05$&	$0.99\pm 0.02$&	$0.80\pm 0.19$&	$  0.99\pm	0.02  $\\
HIV PBS&	$188$&	$0.92\pm 0.02$&	$1.00\pm 0.01$&	$1.00\pm 0.01$&	$1.00\pm 0.02$&	$0.99\pm 0.03$&	$0.91\pm 0.11$&	$  1.00\pm	0.01  $\\
Intron gpII&	$181$&	$0.46\pm 0.13$&	$0.64\pm 0.17$&	$0.64\pm 0.17$&	$0.63\pm 0.17$&	$0.50\pm 0.28$&	$0.49\pm 0.18$&$0.65\pm	0.15$\\
IRES HCV&	$764$&	$0.65\pm 0.11$&	$0.88\pm 0.16$&	$0.45\pm 0.19$&	$0.86\pm 0.17$&	$0.68\pm 0.38$&	$0.85\pm 0.08$&	$  0.88\pm	0.08  $\\
IRES Picorna&	$181$&	$0.84\pm 0.07$&	$0.97\pm 0.03$&	$0.61\pm 0.04$&	$0.96\pm 0.04$&	$0.95\pm 0.04$&	$0.85\pm 0.11$& $0.96\pm	0.04$\\
K chan RES&	$124$&	$0.74\pm 0.10$&	$0.99\pm 0.02$&	$0.98\pm 0.05$&	$0.89\pm 0.19$&	$0.95\pm 0.08$&	$0.58\pm 0.26$&	$  0.95\pm	0.11  $\\
Lysine&	$80$&	$0.50\pm 0.13$&	$0.72\pm 0.13$&	$0.54\pm 0.15$&	$0.71\pm 0.18$&	$0.66\pm 0.16$&	$0.50\pm 0.16$&	$  0.72\pm	0.15 $\\
Retroviral psi&	$89$&	$0.88\pm 0.03$&	$0.93\pm 0.03$&	$0.93\pm 0.03$&	$0.93\pm 0.03$&	$0.92\pm 0.04$&	$0.74\pm 0.12$&	$  0.93	\pm 0.04 $\\
S box&	$91$&	$0.60\pm 0.10$&	$0.75\pm 0.13$&	$0.76\pm 0.16$&	$0.79\pm 0.14$&	$0.67\pm 0.24$&	$0.54\pm 0.16$&	$  0.77\pm	0.12 $\\
SECIS&	$114$&	$0.44\pm 0.16$&	$0.59\pm 0.21$&	$0.62\pm 0.21$&	$0.57\pm 0.25$&	$0.54\pm 0.25$&	$0.39\pm 0.24$&	$ 0.61\pm	0.20   $\\
sno 14q I II&	$44$&	$0.75\pm 0.10$&	$0.92\pm 0.10$&	$0.89\pm 0.16$&	$0.85\pm 0.20$&	$0.89\pm 0.19$&	$0.58\pm 0.27$&	$  0.91\pm	0.13  $\\
SRP bact&	$114$&	$0.48\pm 0.16$&	$0.65\pm 0.21$&	$0.66\pm 0.21$&	$0.63\pm 0.25$&	$0.65\pm 0.21$&	$0.51\pm 0.22$&	$  0.61\pm	0.25  $\\
SRP euk arch&	$122$&	$0.51\pm 0.20$&	$0.62\pm 0.29$&	$0.35\pm 0.17$&	$0.64\pm 0.28$&	$0.64\pm 0.26$&	$0.50\pm 0.26$&	$  0.61\pm	0.29  $\\
T-box&	$18$&	$0.68\pm 0.15$&	$0.77\pm 0.17$&	$0.49\pm 0.17$&	$0.68\pm 0.25$&	$0.70\pm 0.17$&	$0.59\pm 0.21$&	$  0.74\pm	0.15  $\\
TAR&	$286$&	$0.87\pm 0.04$&	$0.99\pm 0.03$&	$0.99\pm 0.02$&	$0.99\pm 0.03$&	$0.98\pm 0.04$&	$0.83\pm 0.19$&	$  0.99\pm	0.04  $\\
THI&	$321$&	$0.45\pm 0.10$&	$0.68\pm 0.16$&	$0.66\pm 0.20$&	$0.68\pm 0.18$&	$0.50\pm 0.29$&	$0.48\pm 0.18$&	$   0.65\pm	0.20 $\\
tRNA&	$2039$&	$0.43\pm 0.12$&	$0.75\pm 0.21$&	$0.85\pm 0.16$&	$0.82\pm 0.19$&	$0.76\pm 0.27$&	$0.66\pm 0.23$&	$  0.72\pm	0.22  $\\
U1&	$82$&	$0.63\pm 0.17$&	$0.79\pm 0.17$&	$0.70\pm 0.13$&	$0.79\pm 0.19$&	$0.80\pm 0.14$&	$0.67\pm 0.20$&	$  0.77\pm	0.17  $\\
U2&	$112$&	$0.64\pm 0.16$&	$0.75\pm 0.17$&	$0.63\pm 0.13$&	$0.76\pm 0.19$&	$0.73\pm 0.22$&	$0.59\pm 0.19$&	$  0.75\pm	0.18  $\\
U6&	$30$&	$0.83\pm 0.06$&	$0.93\pm 0.05$&	$0.89\pm 0.09$&	$0.90\pm 0.08$&	$0.88\pm 0.10$&	$0.72\pm 0.14$&	$ 0.93\pm	0.06   $\\
UnaL2&	$138$&	$0.77\pm 0.08$&	$0.93\pm 0.08$&	$0.92\pm 0.09$&	$0.89\pm 0.15$&	$0.91\pm 0.10$&	$0.65\pm 0.29$&	$  0.94\pm	0.08  $\\
yybP-ykoY&	$127$&	$0.39\pm 0.14$&	$0.58\pm 0.20$&	$0.54\pm 0.23$&	$0.57\pm 0.25$&	$0.40\pm 0.33$&	$0.46\pm 0.22$&	$ 0.56\pm	0.20   $\\
\hline 
Pooled Average&	$249.33$&	$0.63$&	$0.84$&	$0.81$&	$0.84$&	$0.8$&	$0.68$& $0.82$\\
\hline
\end{tabular}}
\smallskip
\caption{Average F1 scores ($\pm$ one standard deviation) for
{\em pairwise global alignment} of {\tt RNAmountAlign} and four widely used
RNA sequence/structure alignment algorithms on the benchmarking set of
8976 pairwise alignments from the
{\tt BRaliBase K2} database \cite{Gardner.nar05}. 
For each indicated Rfam family, the
the number of alignments (NumAln), sequence identity (SeqId), and
F1-scores  for
{\tt RNAmountAlign}, {\tt LocARNA}, {\tt LARA}, {\tt FOLDALIGN}, and
{\tt DYNALIGN} are listed, along with
pooled averages over all 8976 pairwise
alignments.  Parameters used in Eq~(\ref{eqn:similarityMeasure})
for {\tt RNAmountAlign} were similarity matrix RIBOSUM85-60, structural
similarity weight $\gamma=1/2$, gap initiation $g_i=-3$, gap
extension $g_e=-1$.
}
\label{table:F1measures}
\end{table*}

\begin{table*}[]
\centering
\resizebox*{\textwidth}{0.4\textheight}{
\begin{tabular}{|l|cccccccccc|}
\hline
TYPE           & SEED(LENGTH) & MA(LENGTH)   & MA(PPV)   & MA(TIME)  & FA(LENGTH)    & FA(PPV)   & FA(TIME)        & LOC(LENGTH) & LOC(PPV) & LOC(TIME)     \\ \hline
5 8S rRNA      & 158.48$\pm$7.40  & 71.20$\pm$41.55  & 0.80$\pm$0.32 & 3.70$\pm$0.43 & 168.33$\pm$89.23  & 0.75$\pm$0.25 & 509.56$\pm$411.83   & 767.67$\pm$43.35    & 0.01$\pm$0.03    & 9571.39$\pm$6152.56   \\
5S rRNA        & 120.87$\pm$2.09  & 34.79$\pm$25.44  & 0.45$\pm$0.46 & 1.90$\pm$0.13 & 133.81$\pm$84.46  & 0.65$\pm$0.34 & 331.86$\pm$488.57   & 584.00$\pm$23.69    & 0.02$\pm$0.04    & 3093.17$\pm$1934.60   \\
Cobalamin      & 221.03$\pm$13.67 & 28.60$\pm$16.77  & 0.57$\pm$0.44 & 7.67$\pm$1.14 & 451.73$\pm$256.29 & 0.22$\pm$0.28 & 6830.15$\pm$9052.56 & 1028.20$\pm$59.27   & 0.02$\pm$0.02    & 25712.40$\pm$15252.51 \\
Hammerhead 3   & 64.24$\pm$11.08  & 31.88$\pm$20.40  & 0.38$\pm$0.42 & 0.38$\pm$0.11 & 36.91$\pm$31.83   & 0.30$\pm$0.41 & 23.95$\pm$11.81     & 279.05$\pm$38.70    & 0.04$\pm$0.06    & 159.87$\pm$123.44     \\
let-7          & 85.73$\pm$3.11   & 55.37$\pm$28.14  & 0.75$\pm$0.22 & 0.89$\pm$0.10 & 72.95$\pm$27.35   & 0.48$\pm$0.33 & 65.51$\pm$28.66     & 390.76$\pm$21.37    & 0.04$\pm$0.05    & 462.12$\pm$283.01     \\
Lysin          & 193.91$\pm$13.07 & 68.71$\pm$42.73  & 0.30$\pm$0.33 & 6.27$\pm$0.80 & 163.76$\pm$104.21 & 0.57$\pm$0.30 & 554.25$\pm$730.12   & 918.41$\pm$48.19    & 0.03$\pm$0.04    & 18690.26$\pm$10232.32 \\
mir-10         & 75.71$\pm$1.27   & 55.09$\pm$21.97  & 0.67$\pm$0.24 & 0.72$\pm$0.04 & 66.91$\pm$30.83   & 0.48$\pm$0.36 & 45.68$\pm$19.80     & 358.55$\pm$15.96    & 0.03$\pm$0.04    & 333.63$\pm$227.10     \\
Purine         & 102.01$\pm$0.93  & 129.05$\pm$86.84 & 0.41$\pm$0.39 & 1.37$\pm$0.07 & 69.80$\pm$6.70    & 0.88$\pm$0.15 & 87.27$\pm$30.47     & 497.41$\pm$16.81    & 0.03$\pm$0.05    & 2395.40$\pm$1571.67   \\
RFN element    & 147.23$\pm$13.62 & 44.11$\pm$24.91  & 0.94$\pm$0.11 & 2.83$\pm$0.56 & 114.59$\pm$98.77  & 0.80$\pm$0.24 & 619.68$\pm$1289.50  & 687.71$\pm$62.46    & 0.03$\pm$0.05    & 5893.83$\pm$3827.59   \\
S-box leader   & 120.13$\pm$16.14 & 50.35$\pm$30.00  & 0.57$\pm$0.36 & 1.68$\pm$0.44 & 88.72$\pm$60.79   & 0.79$\pm$0.21 & 190.03$\pm$493.08   & 554.09$\pm$55.21    & 0.03$\pm$0.04    & 2399.58$\pm$1484.64   \\
SECIS          & 68.55$\pm$2.88   & 25.76$\pm$21.34  & 0.05$\pm$0.19 & 0.53$\pm$0.05 & 54.25$\pm$53.42   & 0.16$\pm$0.28 & 51.07$\pm$65.81     & 318.53$\pm$16.40    & 0.02$\pm$0.03    & 279.38$\pm$187.58     \\
SNORD113       & 79.69$\pm$6.10   & 40.03$\pm$23.27  & 0.33$\pm$0.42 & 0.75$\pm$0.07 & 47.63$\pm$30.40   & 0.62$\pm$0.40 & 44.32$\pm$18.12     & 373.69$\pm$21.77    & 0.02$\pm$0.02    & 641.43$\pm$421.62     \\
SRP bact       & 96.20$\pm$9.99   & 30.81$\pm$14.92  & 0.69$\pm$0.41 & 0.99$\pm$0.30 & 105.08$\pm$82.04  & 0.66$\pm$0.32 & 225.15$\pm$336.93   & 423.55$\pm$74.67    & 0.02$\pm$0.04    & 726.66$\pm$659.87     \\
THI element    & 117.20$\pm$11.95 & 33.03$\pm$14.43  & 0.51$\pm$0.45 & 1.62$\pm$0.30 & 84.45$\pm$85.58   & 0.75$\pm$0.31 & 253.89$\pm$352.01   & 535.40$\pm$43.83    & 0.02$\pm$0.02    & 2319.39$\pm$1468.99   \\
tRNA           & 76.05$\pm$5.79   & 37.31$\pm$45.09  & 0.23$\pm$0.40 & 0.70$\pm$0.09 & 62.15$\pm$38.30   & 0.67$\pm$0.40 & 73.45$\pm$78.89     & 360.29$\pm$24.06    & 0.02$\pm$0.04    & 479.15$\pm$265.22     \\
Tymo tRNA-like & 86.25$\pm$1.35   & 41.27$\pm$21.96  & 0.50$\pm$0.39 & 0.79$\pm$0.05 & 78.97$\pm$33.70   & 0.76$\pm$0.21 & 84.70$\pm$55.19     & 409.13$\pm$14.22    & 0.04$\pm$0.05    & 684.12$\pm$411.97     \\
U1             & 167.16$\pm$2.58  & 48.36$\pm$32.73  & 0.69$\pm$0.34 & 4.52$\pm$0.16 & 221.36$\pm$121.42 & 0.61$\pm$0.23 & 1755.35$\pm$1255.41 & 804.19$\pm$24.78    & 0.03$\pm$0.05    & 11142.21$\pm$6902.37  \\
U4             & 163.25$\pm$24.55 & 50.64$\pm$27.53  & 0.42$\pm$0.41 & 3.72$\pm$1.30 & 91.75$\pm$41.17   & 0.79$\pm$0.20 & 263.51$\pm$140.53   & 742.17$\pm$84.30    & 0.02$\pm$0.03    & 9361.29$\pm$5839.12   \\
UnaL2          & 54.25$\pm$0.66   & 48.80$\pm$25.71  & 0.70$\pm$0.40 & 0.36$\pm$0.01 & 36.11$\pm$3.30    & 0.99$\pm$0.04 & 23.05$\pm$8.38      & 263.79$\pm$8.94     & 0.03$\pm$0.06    & 171.59$\pm$104.10     \\
ykoK           & 175.39$\pm$7.32  & 82.05$\pm$58.19  & 0.68$\pm$0.36 & 4.67$\pm$0.45 & 147.55$\pm$69.66  & 0.81$\pm$0.20 & 472.79$\pm$583.01   & 844.27$\pm$31.56    & 0.03$\pm$0.05    & 12019.33$\pm$6178.91  \\
ykoK           & 144.26$\pm$63.44 & 81.06$\pm$54.94  & 0.65$\pm$0.38 & 4.74$\pm$0.45 & 144.26$\pm$63.44  & 0.81$\pm$0.20 & 449.03$\pm$526.67   & 482.97$\pm$27.04    & 0.00$\pm$0.00    & 12693.37$\pm$7330.66  \\ \hline
Pooled Average & 118.67$\pm$47.86 & 50.35$\pm$42.33  & 0.53$\pm$0.42 & 2.30$\pm$2.12 & 114.86$\pm$125.33 & 0.64$\pm$0.36 & 625.53$\pm$2554.61  & 556.82$\pm$227.00   & 0.03$\pm$0.04    & 5317.96$\pm$8585.19   \\ \hline
\end{tabular}}
\smallskip
\caption{Comparison of alignment length and positive predictive value (PPV) for
{\em pairwise local alignment} by {\tt RNAmountAlign} against the widely used local alignment
software {\tt FOLDALIGN} and {\tt LocARNA}. Local alignment benchmarking was
performed on  1500
pairwise alignments (75 alignments per family, 20 Rfam families)
extracted from  the Rfam 12.0 database \cite{Nawrocki.nar15}, 
and prepared in a manner analogous to
that of the dataset used in benchmarking {\em multiple} local alignment in \cite{Tabei.b09} -- see text for details.
Parameters used in Eq~(\ref{eqn:similarityMeasure}) of the main text
for {\tt RNAmountAlign} were 
structural similarity weight $\gamma=1/2$, gap initiation $g_i=-3$, gap
extension $g_e=-1$; since reference alignments were required to have at
most 70\% sequence identity, nucleotide similarity matrix RIBOSUM8570-25
was used in {\tt RNAmountAlign}.}
\label{table:localAlignmentBenchmarking}
\end{table*}

Taken together, these results suggest that
{\tt RNAmountAlign} has comparable accuracy, but much faster run time,
hence making it a potentially useful tool for genome scanning applications.
Here it should be stressed
that all benchmarking results used equally weighted contributions of sequence
and ensemble structural similarity; i.e. parameter $\gamma=1/2$ when
computing similarity by Eq~(\ref{eqn:similarityMeasure}).
By setting $\gamma=1$, {\tt RNAmountAlign} alignments depend wholly on
structural similarity (see Figure~\ref{fig:alignmentTRNAwith28percentSeqId}).
Indeed, for the following {\tt BRAliBase 2.1} alignment with 
28\% sequence identity, by setting $\gamma=1$, {\tt RNAmountAlign}
returns the correct alignment.
\begin{quote}
\begin{small}
\begin{verbatim}
GGGGAUGUAGCUCAGUGGUAGAGCGCAUGCUUCGCAUGUAUGAGGCCCCGGGUUCGAUCCCCGGCAUCUCCA
GUUUCAUGAGUAUAGC---AGUACAUUCGGCUUCCAACCGAAAGGUUUUUGUAAACAACCAAAAAUGAAAUA
\end{verbatim}
\end{small}
\end{quote}
of 72 nt tRNA AL671879.2 with 69 nt tRNA D16387.1.
Fig~\ref{fig:alignmentTRNAwith28percentSeqId} shows the superimposed mountain
heights for this alignment.

\subsection{Statistics for pairwise alignment}
\begin{figure*}
\centering
\includegraphics[width=0.8\textwidth]{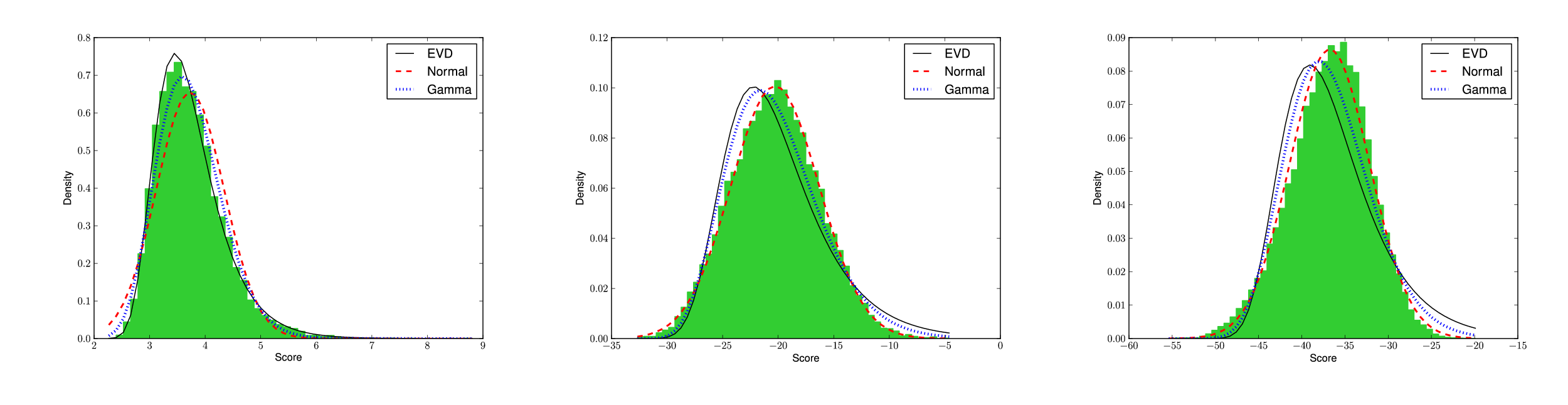}
\caption{Fits of 30-bin relative frequency histograms of scores for
\textit{local} (left), \textit{semiglobal} (middle) and \textit{global} (right) alignments produced by
{\tt RNAmountAlign} for the randomly chosen 5S rRNA 
AY544430.1:375-465 from Rfam 12.0 database 
having A,C,G,U relative frequency
of $0.25,0.27,0.26,0.21$. A total of 10,000 random sequences having identical
expected nucleotide relative frequencies were generated, each of length 400 nt for local/semiglobal and 100 nt for global. Local (left), semiglobal (middle) and global (right)  alignments were computed by
{\tt RNAmountAlign}, in each case fitting the data with the normal (ND),
extreme value (EVD) and gamma (GD) distributions. As expected by 
Karlin-Altschul theory \cite{Karlin.pnas90}, local alignment scores are
best fit by EVD, while semiglobal alignment scores are best fit
by ND (results supported by data not shown, involving
computations of variation distance, 
symmetrized Kullback-Leibler distance, and $\chi^2$ goodness-of-fit tests).
}
\label{fig:distributionFitForSemiglobal}
\end{figure*}

\begin{figure}
\centering
\includegraphics[width=0.8\textwidth]{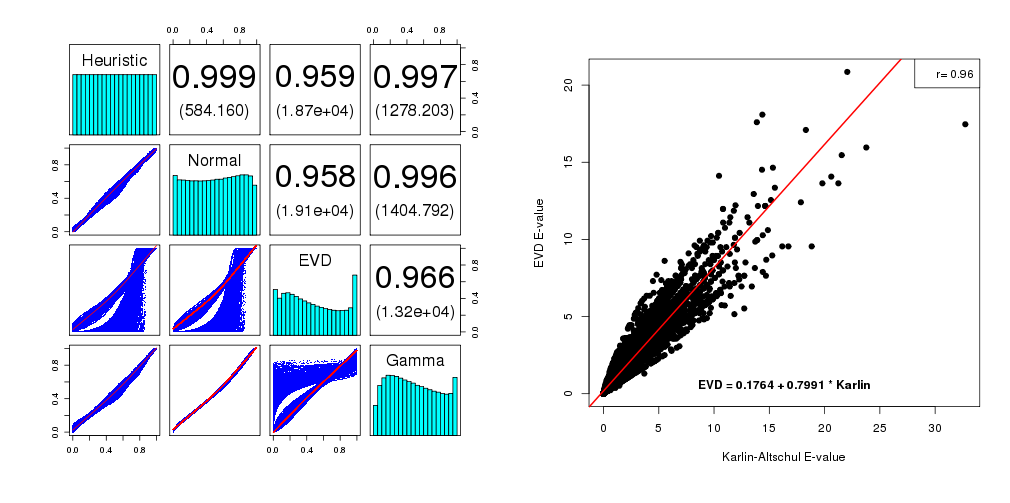}
\caption{
(\textbf{Left})Pearson correlation values and scatter plots for $p$-values of
\textit{semiglobal alignment}(query search) scores between Rfam sequences and
random RNA. 
For each score in a set of
2.5 million global pairwise alignment scores, a $p$-value was computed
by direct counts (heuristic), or by data fitting the normal (ND), extreme value (EVD), or gamma (GD) distributions.
Pairwise
Pearson correlation values were computed and displayed in the upper triangular
portion of the figure, with sums of squared residuals shown in parentheses, and
histograms of $p$-values along the diagonal.  
It follows that ND $p$-values correlate best with
heuristic $p$-values, where the latter is assumed to be the gold standard.
(\textbf{Right})Scatter plot of expect values $E_{\mbox{\tiny ML}}$, computed by maximum likelihood, following the
method described in \cite{eddy:RSEARCH} ($y$-axis) and expect values $E_{\mbox{\tiny KA}}$, computed
by our implementation of the Karlin-Altschul, as described in the text.
The regression equation is 
$E_{\mbox{\tiny ML}} = 0.1764 + 0.7991 \cdot E_{\mbox{\tiny KA}}$; Pearson correlation between 
$E_{\mbox{\tiny ML}}$ and $E_{\mbox{\tiny KA}}$ is 96\%, with correlation $p$-value of 
$2 \cdot 10^{-16}$. Expect values were determined from local alignment scores computed by the
genome scanning form of {\tt RNAmountAlign} with query tRNA AB031215.1/9125-9195 and targets
consisting of 300 nt windows (with 200 nt overlap) from
{\em E. coli str. K-12} substr. MG1655 with GenBank accession code AKVX01000001.1. From the
tRNA query sequence, the values $p_A,p_C,p_G,p_U$ for
nucleotide relative frequencies, are determined, then average base pairing probabilities
$p_{\op},p_{\bullet},p_{\cp}$ are computed by {\tt RNAfold -p} \cite{Lorenz.amb11}.
For the current 300 nt target window, the nucleotide relative frequencies
$p'_A,p'_C,p'_G,p'_U$ are computed, then precomputed probabilities
$p'_{\op},p'_{\bullet},p'_{\cp}$ are obtained from SI Table~\ref{table:precomputedProb}.
From these values, scaling factor $\alpha_{\mbox{\tiny seq}}$ and shift $\alpha_{\mbox{\tiny str}}$,
were computed; with structural similarity weight $\gamma=1/2$, the overall similarity function
from Eq~(\ref{eqn:similarityMeasure}) in the text was determined.
}
\label{fig:correlationPvaluesGlobalSemiglobalAlignment}
\end{figure}

Fig~\ref{fig:distributionFitForSemiglobal} shows fits of the 
relative frequency histogram of alignment scores with
the normal (ND), extreme value (EVD) and gamma (GD) distributions, where local
[resp. semiglobal] alignment scores are shown in the left [resp.
right] panel. The EVD provides the best fit for local alignment 
sequence-structure similarity scores, as expected by Karlin-Altschul theo \cite{karlinDemboKawabata,Karlin.pnas90}. Moreover, Fig \ref{fig:correlationPvaluesGlobalSemiglobalAlignment} shows a 96\% correlation between (expect) E-values computed by
our implementation of the Karlin-Altschul method, and E-values obtained by
maximum likelihood fitting of local alignment scores. In contrast, the ND provides the best fit for semiglobal sequence/structure alignment similarity scores, at
least for the sequence considered in
Fig~\ref{fig:distributionFitForSemiglobal}. This is not an isolated
phenomenon, as shown in 
Fig~\ref{fig:correlationPvaluesGlobalSemiglobalAlignment}, which
depicts scatter plots, Pearson correlation values and sums of squared
residuals (SSRs) when computing $p$-values for 
semiglobal (query search) alignment scores between Rfam sequences and
random RNA.  
As explained earlier, a pool of 2220 sequences from the Rfam 12.0 database \cite{Nawrocki.nar15} was created by selecting one sequence of length at most $200$ nt from each family, with the property that base pair distance
between its minimum free energy (MFE) structure and the Rfam consensus
structure was a minimum. Then 500 sequences were randomly selected from
this pool, and for each of five gap
initiation and extension costs $g_i = -5,-4,-3,-2,-1$ with
$g_e = \frac{g_i}{3}$. Taking each of the 500 sequences successively as
query sequence and for each choice of
parameters, $1000$ random 400 nt RNAs were generated with
the same expected nucleotide relative frequency as that of the query.
For each
alignment score $z$ for query and random target, the $p$-value was computed as
$1$ minus the cumulative density function, $1-CDF(z)$, for 
fitted normal (ND), extreme value (EVD) and gamma (GD) distributions, 
thus defining $1000$ $p$-values.
Additionally, a heuristic $p$-value was determined by
calculating the proportion of alignment scores for given query that exceed
$z$.  For each set of
$2.5$ million ($500 \times 5 \times 1000$) $p$-values (heuristic, ND, EVD, GD),
Pearson correlation values were computed and displayed in the upper triangular
portion of 
Fig~\ref{fig:correlationPvaluesGlobalSemiglobalAlignment}, 
with SSRs shown in parentheses. Note that
residuals were computed for regression equation
$\text{row} = m \cdot \text{column} + b$, where column values constitute the
independent variable. 
Assuming that heuristic $p$-values constitute the reference standard,
it follows that $p$-values computed from the normal distribution
correlate best with semiglobal alignment
scores computed by {\tt RNAmountAlign}.

Earlier studies have suggested that
protein global alignment similarity scores using PAM120, PAM250, BLOSUM50, and
BLOSUM62 matrices appear to be fit best by the gamma distribution (GD) 
\cite{Pang.bb05}, and that semiglobal RNA sequence alignment
similarity scores 
(with no contribution from structure) appear to be best fit by GD 
\cite{Hertel.nar09}.  However, in our preliminary studies
(not shown), it appears that the type of distribution (ND, EVD, GD) that
best fits {\tt RNAmountAlign} semiglobal alignment
depends on the gap costs applied (indeed, for
certain choices, EVD provides the best fit).
Since there is no mathematical theory concerning alignment score distribution
for global or semiglobal alignments, it must be up to the user to decide
which distribution provides the most reasonable $p$-values.


\subsection{Multiple alignment}
\begin{figure*}
\centering
\includegraphics[width=0.8\textwidth]{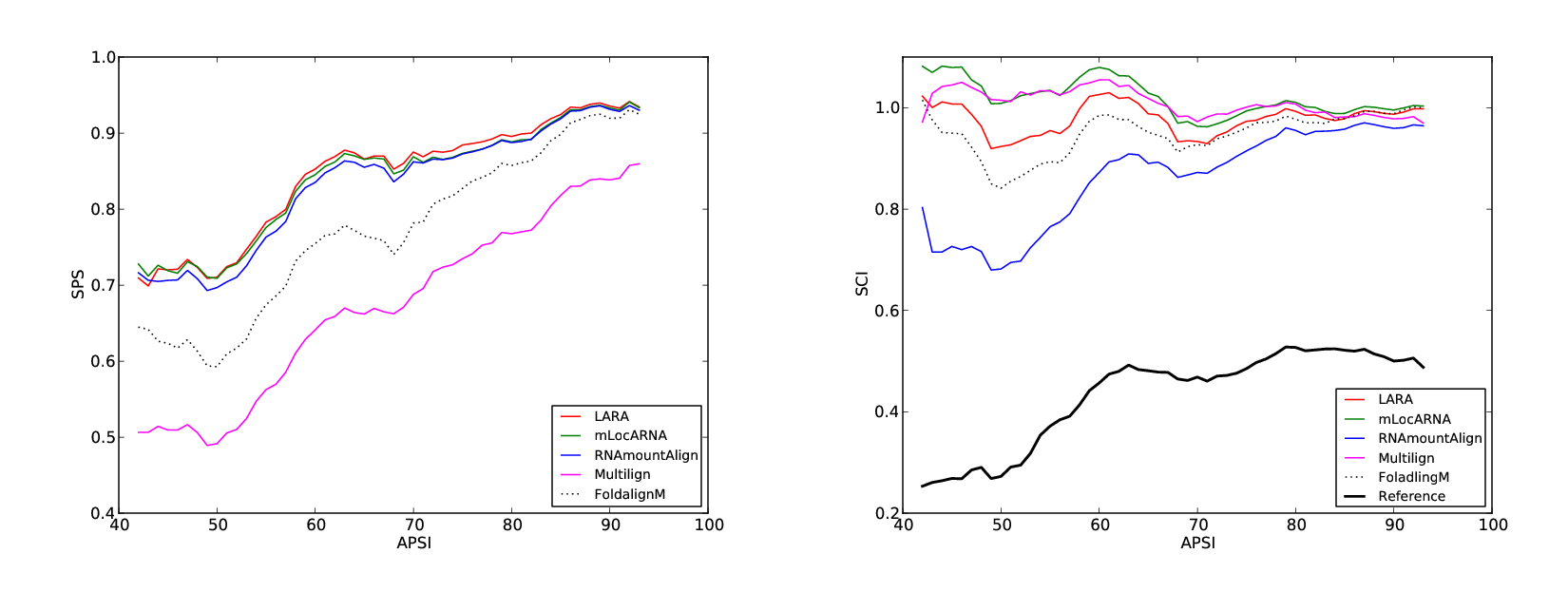}

\caption{Sum-of-pairs(SPS) score (left) and structural conservation index (SCI) (right) for \textit{multiple global alignments} using 
{\tt RNAmountAlign}, {\tt LARA}, {\tt mLocARNA}, {\tt FoldalignM} and {\tt Multilign} . 
SPS and SCI are shown as a function of  
average pairwise sequence identity(APSI) in the
k5 {\tt BRAliBase 3} database used for benchmarking.
}
\label{fig:multiFSCI}
\end{figure*}

\begin{figure*}
\centering
\includegraphics[width=0.8\textwidth]{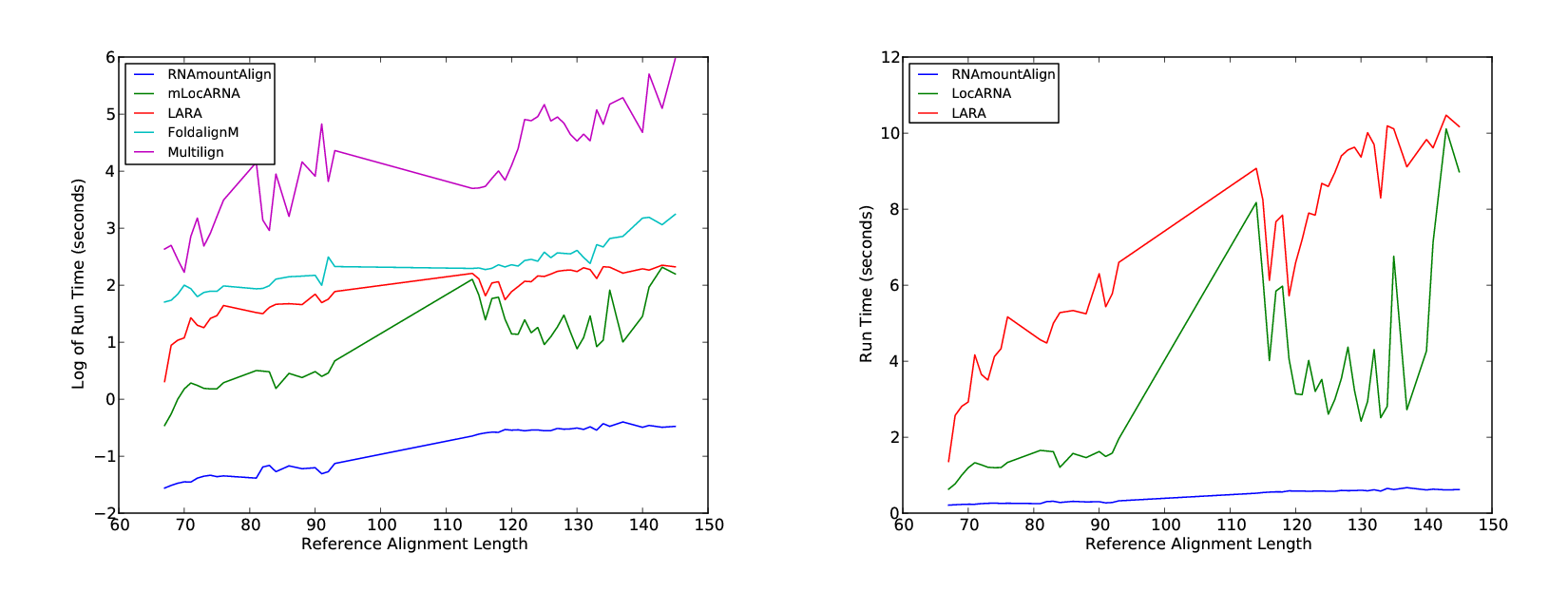}
\caption{Run time of \textit{multiple global alignment} for {\tt RNAmountAlign}, {\tt mLocARNA} and {\tt LARA}, {\tt FoldalignM} and {\tt Multilign}. (Left) Log run time is as shown a function of reference alignment length for K5 alignments in {\tt Bralibase 3}. (Right) Actual run time in seconds for {\tt mLocARNA} and {\tt LARA}.
}
\label{fig:runTimeMulti}
\end{figure*}

We benchmarked {\tt RNAmountAlign} with the software {\tt LARA}, 
{\tt mLocARNA}, {\tt FOLDALIGNM} and {\tt Multilign} for 
\textit{multiple global} K5 alignments in {\tt Bralibase 3}. 
{\tt STRAL} is not included since the source code could not be compiled.
Fig \ref{fig:multiFSCI} indicates average SPS and SCI as a function of 
average pairwise sequence identity (APSI). We used the {\tt -sci} flag of 
{\tt RNAalifold} to compute SCI from the output of each software without
reference to the reference alignment. Fig ~\ref{fig:multiFSCI} indicates that  
SCI values for outputs from various alignment algorithms is higher
than the SCI value from reference alignments,
suggesting that the consensus structure obtained from 
sequence/structure alignment algorithms has a larger number of 
base pairs than the the consensus structure obtained from reference 
alignments (this phenomenon was also in \cite{Smith2017}).
Fig ~\ref{fig:multiFSCI} indicates that {\tt RNAmountAlign} produces 
SPS scores comparable to {\tt mLocARNA} and {\tt LARA} and higher than 
{\tt Multilign} and {\tt FOLDALIGNM} while the SCI score obtained from 
{\tt RNAmountAlign} are slightly lower than other software. Averaging over all sequences, the SPS
scores for {\tt RNAmountAlign}, {\tt LARA}, 
{\tt mLocARNA}, {\tt FOLDALIGNM} and {\tt Multilign} were respectively: $0.84 \pm 0.17$, $0.85 \pm 0.17$, $0.84 \pm 0.17$, $0.77 \pm 0.22$, and $0.84 \pm 0.19$. The left panel of
Fig \ref{fig:runTimeMulti} indicates the run time of all software on a
logarithmic scale, while the right panel shows the actual run time in
seconds for {\tt RNAmountAlign} as well as that of the next two fastest 
algorithms, {\tt mLocARNA} and {\tt LARA}.
This figure clearly shows that {\tt RNAmountAlign} has much faster run time 
than all other software in our benchmarking tests, thus confirming the 
earlier result from pairwise benchmarking.

\section{Conclusion}

{\tt RNAmountAlign} is a new {\tt C++} software package for
RNA local, global, and semiglobal sequence/structure alignment, which 
provides accuracy comparable with that of a number of widely used programs,
but provides much faster run time. {\tt RNAmountAlign} additionally computes 
E-values for local alignments, using Karlin-Altschul statistics, as well as
$p$-values for normal, extreme value and gamma distributions by parameter
fitting.

\section{Acknowledgements}

Research supported by National Science Foundation grant DBI-1262439. 
Any opinions, findings, and conclusions or recommendations
expressed in this material are those of the authors and do not 
necessarily reflect the views of the National Science Foundation.

\bibliographystyle{plain}

\makeatletter
\renewcommand{\@biblabel}[1]{\quad#1.}
\makeatother

\hfill\clearpage


\hfill\clearpage

\appendix
\section{Supplementary Information}

\subsection{Software usage}
{\tt RNAmountAlign} performs local, semiglobal, and global sequence/structure alignments. By default the global alignment is computed unless flags {\tt -local} or {\tt -semi} are used to perform local and semiglobal alignments, respectively. In the simplest case, the program could be run with
\begin{lstlisting}
> ./RNAmountAlign -f <inputFasta>
or
> ./RNAmountAlign -s seq1 seq2
\end{lstlisting}
The parameters that were used to produce the results in the main text are used as the default by the software: structural similarity weight $\gamma=0.5$, gap initiation $g_i=-3$, and gap extension $g_e=-1$. The weight factor $\gamma$ defines the importance of structural similarity versus sequence similarity. When $\gamma=0$ only sequence similarity is considered, while $\gamma=1$ only uses the incremental ensemble mountain heights for the alignment. As an example, let's consider the following two toy sequences each forming a stem loop secondary structure
\begin{lstlisting}
>seq1
AAAAAAAAAACCCCCUUUUUUUUUU
((((((((((.....)))))))))) (-2.1)
>seq2
CCCCCCCAAAAGGGGGGG
(((((((....))))))) (-15.7)
\end{lstlisting}
Running the software considering only sequence similarity with gap initiation and extension penalties of -2 and -1, respectively, by the command
\begin{tiny}
\begin{lstlisting}
> ./RNAmountAlign -s AAAAAAAAAACCCCCUUUUUUUUUU CCCCCCCAAAAGGGGGGG -gamma 0 -gi -2 -ge -1
\end{lstlisting}
\end{tiny}
produces the following alignment
\begin{lstlisting}
seq1	1	AAAAAAAAAACCCCCUUUUUUUUUU  25  
seq2	1	-------CCCCCCCAAAAGGGGGGG  18
\end{lstlisting}
where four {\tt C} nucleotides are aligned together,
regardless of the fact that in the secondary structure for the first sequence, 
they are found in an apical loop region,
while in the secondary structure for the second sequence, 
they are part of a stem. However, using {\tt -gamma 1} returns 
\begin{lstlisting}
seq1	1	AAAAAAAAAACCCCCUUUUUUUUUU  25   
seq2	1	CCCCCCC----AAAAGGGGGGG---  18  
\end{lstlisting}
where the opening, closing and unpaired bases are aligned to each other. Finally, using {\tt -gamma 0.5} gives
\begin{lstlisting}
seq1	1	AAAAAAAAAACCCCCUUUUUUUUUU  25   
seq2	1	CCCCCCCAAAA-------GGGGGGG  18  
\end{lstlisting}
where both sequence and structural similarity are equally weighted.
The default nucleotide similarity matrix is RIBOSUM85-60. 
Other RIBOSUM matrices are included in the software and can be selected 
with {\tt -m} flag based on the user's knowledge of 
divergence of the input sequences.\\
{\tt RNAmountAlign} computes the consensus secondary structure by calling {\tt alifold()} function from {\tt libRNA.a} in the Vienna RNA Package when flag {\tt -alifold} is used. For example the following command outputs the consensus structure in addition to the alignment for the same sequences indicated in Fig \ref{fig:alignmentTRNAwith28percentSeqId} of the main text. See Fig \ref{fig:consStr}.\\
\begin{lstlisting}
> ./RNAmountAlign -f examples/trna.fa -alifold -global
\end{lstlisting}

Computation of alignment statistics depends on the alignment type. As discussed in the main text, local alignment scores follow extreme value distribution(EVD) while global and semiglobal scores tend to follow normal distribution(ND). Flag {\tt -stat} can be set to compute both $E$-values and $p$-values, where the transformation between $E$-values and $p$-values is made by $p=1-exp(-E)$. For global and semiglobal alignments, the first (query) sequence is aligned to a number of random RNAs, defined by {\tt -num} flag, with the same nucleotide composition as the second sequence (target), then the random alignment scores are fitted to normal distribution and a $p$-value is returned.
\begin{lstlisting}
> ./RNAmountAlign -f examples/trna.fa -global -stat -num 100
\end{lstlisting} 
As part of the output, $p$-value from ND normal fitting of 100 random alignment scores is reported:
\begin{lstlisting} 
Normal distribution E-value: 0.0476148
Normal distribution p-value: 0.046499
\end{lstlisting} 
For local alignments either Karlin-Altschul statistics (default) or EVD fitting can be computed. Let's consider an example of a local alignment between two purine riboswitches with Rfam seed alignment length of 102 and sequence identity $0.58$. Random flanking regions with the same nucleotide composition are added to the seed alignment as discussed in the main text to obtain two sequences of length $408$ and $400$. The local alignment between these two sequences has length $53$ with extremely low $E$-value, with the property that all pairs in the local alignment are found in the reference seed alignment ($PPV=1$). $E$-value from Karlin-Altschul statistics can be obtained very fast from the following command: 
\begin{lstlisting}
> ./RNAmountAlign -f examples/RF00167_1.raw -local -stat
Karlin-Altschul E-value: 2.52137e-06
Karlin-Altschul p-value: 2.52137e-06
\end{lstlisting}
Computation of $E$-value from EVD fitting is more accurate but slower:
\begin{lstlisting}
> ./RNAmountAlign -f examples/RF00167_1.raw -local -stat -evd -num 200 
Extreme value distribution E-value: 4.41417e-05
Extreme value distribution P-value: 4.41408e-05
\end{lstlisting}
{\tt RNAmountAlign} computes Karlin-Altschul $E$-values from maximum likelihood method described in the main text, and then multiplies it by the regression coefficient of $0.7991$, indicated in the right panel of Fig \ref{fig:correlationPvaluesGlobalSemiglobalAlignment}, to obtain an estimated $E$-value. Therefore, there might be discrepancy between the EVD fitting and Karlin-Altschul $E$-values. For the most accurate statistics EVD fitting is recommended.\\

Our software could also be used for searching a query sequence defined by {\tt -qf <fastaFile>} in a target sequence defined by {\tt -tf <fastaFile>}. The search computes semiglobal alignments of the query to sliding windows of the target, and returns the aligned segments of the target sorted by $p$-value. The query is aligned to windows of a fixed size defined by {\tt -window}, sliding by steps defined by {\tt  -step} flag. To compute the statistics, random alignment scores are computed and fitted to ND. However, the software does not compute random alignments for each window separately as it would be very slow. Instead, following \cite{eddy:RSEARCH}, the range of the GC-content of the target sequence over all the sliding windows is first obtained and binned using bin size defined by {\tt -gc}. For each GC-content bin, fitting paremeters are precomputed by generating a number of random sequences whose GC-content is equal to the bin midpoint, aligning the query to random sequences, and fitting random alignment scores to normal distribution. For each sliding window the corresponding precomputed parameters are used for the computation of $p$-value. As an example, a random tRNA from Rfam 12.0 whose minimum free energy structure has the minimum base pair distance to the Rfam consensus structure was selected and used as the query to search \textit{E. coli} K12 MG1655 genome using window size $300$ and step size $200$ by the following command. 
\begin{lstlisting}
> ./RNAsearch -qf examples/tRNAscan.fa -tf examples/ecoli_MG1655.fa -window 300 -step 200 -gc 10 -num 1000
\end{lstlisting}
The output contains:
\begin{lstlisting}
GC Bins: [0.23-0.33),[0.33-0.43),[0.43-0.53),[0.53-0.63),[0.63-0.73),[0.73-0.74]
1000 random seqs of size 300 generated for each each GC bin.

Fitting to Normal:
\end{lstlisting}
\texttt{
\begin{tabular}{lll}
GC\_Content & Location\_Param & Scale\_Param \\
0.283       & -12.18          & 1.96         \\
0.383       & -13.41          & 2.03         \\
0.483       & -15.01          & 2.05         \\
0.583       & -16.84          & 2.05         \\
0.683       & -18.98          & 2.16         \\
0.735       & -20.08          & 2.06   
\end{tabular}}\\
\medskip
As indicated, six GC bins are generate in range $[0.23-0.74]$; for each bin 1000 random sequences whose GC-content are equal to the average GC-content of the bins are generated, aligned to the query and their fitted location (mean) and scale (standard deviation) parameters are precomputed to be used for computation of $p$-values. From the top 20 hits of our software, the first 18 are reported to be tRNAs by {\tt tRNAscan-SE} \cite{Lowe.nar16}.\\

To see all the full parameter list for the software please use
\begin{lstlisting}
> ./RNAmountAlign -h
\end{lstlisting}

\renewcommand{\thefigure}{S\arabic{figure}}
\setcounter{figure}{0}

\renewcommand{\thetable}{S\arabic{table}}
\setcounter{table}{0}

\begin{figure*}
\centering
\includegraphics[width=0.6\textwidth]{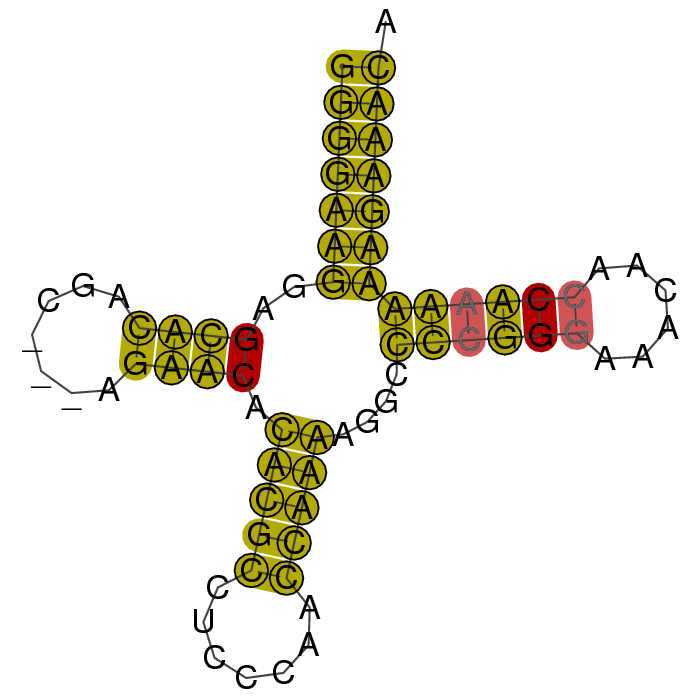}
\caption{Consensus structure for the pairwise alignment indicated in Fig \ref{fig:alignmentTRNAwith28percentSeqId} of the main text. The consensus structure is computed by a calling function {\tt alifold()} from Vienna RNA Package. The figure is obtained from {\tt RNAalifold} web server.}
\label{fig:consStr}
\end{figure*}

\begin{figure*}
\centering
\includegraphics[width=0.8\textwidth]{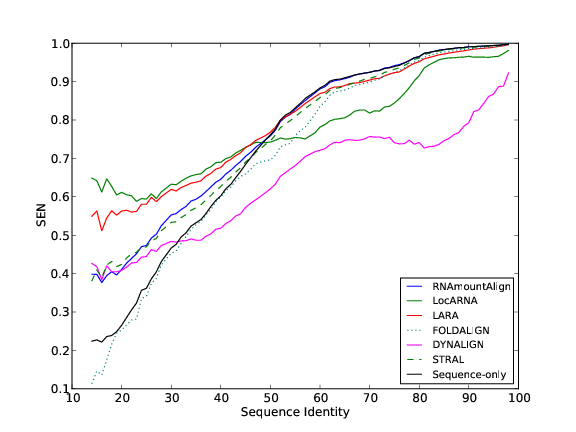}
\caption{Average sensitivity for
{\tt RNAmountAlign}, {\tt LocARNA}, {\tt LARA}, {\tt FOLDALIGN}, {\tt DYNALIGN}, {\tt STRAL}
 and sequence-only alignments ($\gamma=0$) for \textit{pairwise global alignment}. 
Sensitivity is shown as a function of sequence identity for pairwise alignments in the
{\tt BRAliBase 2.1} database used for benchmarking.
}
\label{fig:sensglobalAlignmentRfam}
\end{figure*}

\begin{figure*}
\centering
\includegraphics[width=0.8\textwidth]{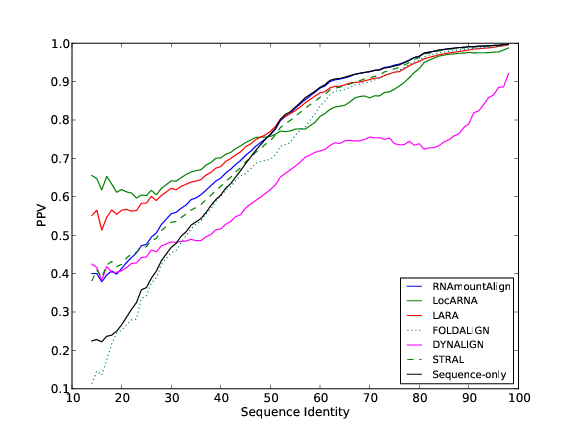}
\caption{Average positive predictive value (PPV) for
{\tt RNAmountAlign}, {\tt LocARNA}, {\tt LARA}, {\tt FOLDALIGN}, {\tt DYNALIGN}, {\tt STRAL}
 and sequence-only alignments($\gamma=0$) for \textit{pairwise global alignment}. 
PPV is shown as a function of sequence identity for pairwise alignments in the
{\tt BRAliBase 2.1} database used for benchmarking.
}
\label{fig:PPVglobalAlignmentRfam}
\end{figure*}

\begin{table*}
\centering
\begin{tiny}
\begin{tabular}{|l|cccccccc|}
\hline
Type           & NumAln & SeqId     & MA(sen)   & LOC(sen) & LARA(sen) & FA(sen)   & DA(sen)   & STRAL(sen) \\
\hline
5.8 S rRNA     & 76     & 0.90±0.09 & 0.95±0.07 & 0.87±0.14    & 0.89±0.11 & 0.65±0.22 & 0.66±0.22 & 0.71±0.15  \\
5S rRNA        & 1162   & 0.60±0.14 & 0.83±0.17 & 0.87±0.13    & 0.84±0.16 & 0.85±0.14 & 0.69±0.17 & 1.00±0.02  \\
Cobalamin      & 188    & 0.43±0.10 & 0.55±0.16 & 0.30±0.13    & 0.48±0.20 & 0.43±0.24 & 0.37±0.19 & 1.00±0.02  \\
Entero 5 CRE   & 48     & 0.88±0.06 & 0.98±0.05 & 0.99±0.04    & 0.99±0.05 & 0.99±0.02 & 0.87±0.12 & 0.88±0.16  \\
Entero CRE     & 65     & 0.80±0.07 & 1.00±0.00 & 0.99±0.03    & 0.97±0.06 & 0.99±0.03 & 0.77±0.16 & 0.90±0.16  \\
Entero OriR    & 49     & 0.84±0.06 & 0.94±0.07 & 0.91±0.09    & 0.94±0.08 & 0.94±0.07 & 0.84±0.15 & 0.93±0.06  \\
gcvT           & 167    & 0.44±0.13 & 0.59±0.19 & 0.60±0.24    & 0.57±0.25 & 0.40±0.33 & 0.44±0.19 & 0.77±0.17  \\
Hammerhead 1   & 53     & 0.71±0.17 & 0.89±0.13 & 0.90±0.12    & 0.87±0.16 & 0.83±0.25 & 0.53±0.27 & 0.92±0.03  \\
Hammerhead 3   & 126    & 0.66±0.21 & 0.86±0.21 & 0.88±0.21    & 0.88±0.21 & 0.79±0.31 & 0.71±0.31 & 1.00±0.01  \\
HCV SLIV       & 98     & 0.85±0.05 & 0.99±0.03 & 0.98±0.04    & 0.98±0.03 & 0.99±0.03 & 0.81±0.34 & 0.94±0.08  \\
HCV SLVII      & 51     & 0.83±0.09 & 0.97±0.06 & 0.96±0.06    & 0.93±0.10 & 0.95±0.07 & 0.72±0.22 & 0.95±0.07  \\
HepC CRE       & 45     & 0.86±0.06 & 1.00±0.00 & 1.00±0.00    & 1.00±0.00 & 1.00±0.00 & 0.77±0.29 & 0.82±0.20  \\
Histone3       & 84     & 0.78±0.09 & 1.00±0.00 & 1.00±0.00    & 1.00±0.00 & 1.00±0.00 & 1.00±0.00 & 0.95±0.07  \\
HIV FE         & 733    & 0.87±0.04 & 1.00±0.02 & 1.00±0.02    & 0.98±0.05 & 0.99±0.05 & 0.65±0.29 & 0.99±0.03  \\
HIV GSL3       & 786    & 0.86±0.04 & 0.99±0.02 & 0.99±0.02    & 0.98±0.05 & 0.99±0.03 & 0.81±0.19 & 0.88±0.11  \\
HIV PBS        & 188    & 0.92±0.02 & 1.00±0.01 & 1.00±0.01    & 1.00±0.02 & 0.99±0.03 & 0.92±0.10 & 0.61±0.29  \\
Intron gpII    & 181    & 0.46±0.13 & 0.64±0.17 & 0.63±0.17    & 0.62±0.18 & 0.50±0.28 & 0.49±0.18 & 0.61±0.25  \\
IRES HCV       & 764    & 0.65±0.11 & 0.87±0.16 & 0.32±0.14    & 0.85±0.17 & 0.67±0.38 & 0.85±0.08 & 0.97±0.06  \\
IRES Picorna   & 181    & 0.84±0.07 & 0.97±0.03 & 0.45±0.03    & 0.96±0.04 & 0.95±0.04 & 0.85±0.10 & 0.74±0.18  \\
K chan RES     & 124    & 0.74±0.10 & 0.99±0.02 & 0.98±0.05    & 0.90±0.19 & 0.95±0.08 & 0.59±0.26 & 0.96±0.04  \\
Lysine         & 80     & 0.50±0.13 & 0.72±0.13 & 0.44±0.13    & 0.71±0.18 & 0.65±0.16 & 0.50±0.16 & 0.54±0.17  \\
Retroviral psi & 89     & 0.88±0.03 & 0.93±0.03 & 0.93±0.03    & 0.93±0.03 & 0.92±0.04 & 0.74±0.12 & 0.99±0.03  \\
S box          & 91     & 0.60±0.10 & 0.75±0.13 & 0.75±0.17    & 0.79±0.14 & 0.67±0.24 & 0.54±0.16 & 1.00±0.00  \\
SECIS          & 114    & 0.44±0.16 & 0.58±0.21 & 0.62±0.21    & 0.57±0.25 & 0.54±0.25 & 0.39±0.24 & 0.61±0.20  \\
sno 14q I II   & 44     & 0.75±0.10 & 0.92±0.10 & 0.89±0.16    & 0.85±0.20 & 0.89±0.19 & 0.59±0.27 & 0.99±0.02  \\
SRP bact       & 114    & 0.48±0.16 & 0.65±0.21 & 0.65±0.21    & 0.63±0.25 & 0.64±0.21 & 0.52±0.22 & 0.61±0.20  \\
SRP euk arch   & 122    & 0.51±0.20 & 0.62±0.29 & 0.24±0.12    & 0.64±0.29 & 0.64±0.26 & 0.51±0.26 & 0.65±0.20  \\
T-box          & 18     & 0.68±0.15 & 0.77±0.17 & 0.36±0.13    & 0.68±0.25 & 0.70±0.17 & 0.59±0.21 & 1.00±0.00  \\
TAR            & 286    & 0.87±0.04 & 0.99±0.03 & 0.99±0.02    & 0.99±0.03 & 0.98±0.04 & 0.84±0.19 & 0.91±0.13  \\
THI            & 321    & 0.45±0.10 & 0.67±0.16 & 0.65±0.21    & 0.68±0.18 & 0.50±0.29 & 0.48±0.18 & 0.65±0.15  \\
tRNA           & 2039   & 0.43±0.12 & 0.75±0.21 & 0.84±0.16    & 0.81±0.19 & 0.76±0.27 & 0.66±0.23 & 0.77±0.12  \\
U1             & 82     & 0.63±0.17 & 0.78±0.17 & 0.61±0.11    & 0.78±0.19 & 0.80±0.14 & 0.67±0.20 & 0.96±0.10  \\
U2             & 112    & 0.64±0.16 & 0.75±0.17 & 0.51±0.11    & 0.76±0.19 & 0.73±0.22 & 0.60±0.19 & 0.55±0.20  \\
U6             & 30     & 0.83±0.06 & 0.93±0.05 & 0.89±0.09    & 0.90±0.08 & 0.88±0.10 & 0.72±0.14 & 0.74±0.15  \\
UnaL2          & 138    & 0.77±0.08 & 0.93±0.08 & 0.92±0.09    & 0.88±0.15 & 0.91±0.09 & 0.65±0.29 & 0.87±0.08  \\
yybP-ykoY      & 127    & 0.39±0.14 & 0.57±0.21 & 0.51±0.23    & 0.56±0.26 & 0.39±0.33 & 0.46±0.22 & 0.73±0.22  \\
\hline
\hline
Pooled Average & 249.33 & 0.63      & 0.83      & 0.78         & 0.84      & 0.80       & 0.68      & 0.82     \\
\hline
\end{tabular}
\end{tiny}
\caption{Average sensitivity scores ($\pm$ one standard deviation) for
{\em pairwise global alignment} of {\tt RNAmountAlign} and four widely used
RNA sequence/structure alignment algorithms on the benchmarking set of
8976 pairwise alignments from the
{\tt BRaliBase K2} database \cite{Gardner.nar05}.
For each indicated Rfam family, the
the number of alignments (NumAln), sequence identity (SeqId), and
sensitivity scores  for
{\tt RNAmountAlign}, {\tt LocARNA}, {\tt LARA}, {\tt FOLDALIGN}, and
{\tt DYNALIGN} are listed, along with
pooled averages over all 8976 pairwise
alignments.  Parameters used in Eq~(\ref{eqn:similarityMeasure}) 
of the main text
for {\tt RNAmountAlign} were similarity matrix RIBOSUM85-60, structural
similarity weight $\gamma=1/2$, gap initiation $g_i=-3$, gap
extension $g_e=-1$.
}
\label{table:globalSensitivity}
\end{table*}

\begin{table*}									
\centering
\begin{tiny}									
\begin{tabular}{|l|cccccccc|}
\hline 									
Type           & NumAln & SeqId     & MA(ppv)   & LOC(ppv) & LARA(ppv) & FA(ppv)   & DA(ppv)   & STRAL(ppv) \\
\hline
5.8 S rRNA     & 76     & 0.72±0.13 & 0.90±0.09 & 0.82±0.07    & 0.87±0.15 & 0.89±0.11 & 0.66±0.22 & 0.88±0.12  \\
5S rRNA        & 1162   & 0.60±0.14 & 0.84±0.16 & 0.88±0.12    & 0.85±0.16 & 0.86±0.14 & 0.68±0.17 & 0.82±0.20  \\
Cobalamin      & 188    & 0.43±0.10 & 0.56±0.16 & 0.54±0.23    & 0.49±0.20 & 0.43±0.24 & 0.36±0.19 & 0.54±0.17  \\
Entero 5 CRE   & 48     & 0.88±0.06 & 0.98±0.04 & 0.99±0.04    & 0.99±0.05 & 0.99±0.02 & 0.86±0.13 & 0.97±0.06  \\
Entero CRE     & 65     & 0.80±0.07 & 1.00±0.00 & 0.99±0.03    & 0.96±0.08 & 0.99±0.04 & 0.74±0.18 & 0.99±0.03  \\
Entero OriR    & 49     & 0.84±0.06 & 0.95±0.07 & 0.94±0.08    & 0.94±0.08 & 0.94±0.07 & 0.84±0.15 & 0.96±0.08  \\
gcvT           & 167    & 0.44±0.13 & 0.62±0.18 & 0.63±0.23    & 0.58±0.25 & 0.41±0.34 & 0.44±0.19 & 0.62±0.20  \\
Hammerhead 1   & 53     & 0.71±0.17 & 0.90±0.13 & 0.90±0.11    & 0.87±0.16 & 0.83±0.25 & 0.51±0.27 & 0.88±0.16  \\
Hammerhead 3   & 126    & 0.66±0.21 & 0.87±0.20 & 0.88±0.21    & 0.89±0.20 & 0.80±0.30 & 0.71±0.31 & 0.91±0.15  \\
HCV SLIV       & 98     & 0.85±0.05 & 0.99±0.03 & 0.98±0.04    & 0.98±0.03 & 0.99±0.03 & 0.80±0.34 & 0.99±0.03  \\
HCV SLVII      & 51     & 0.83±0.09 & 0.97±0.06 & 0.96±0.06    & 0.93±0.10 & 0.95±0.07 & 0.69±0.22 & 0.95±0.07  \\
HepC CRE       & 45     & 0.86±0.06 & 1.00±0.00 & 1.00±0.00    & 1.00±0.00 & 1.00±0.00 & 0.76±0.29 & 1.00±0.00  \\
Histone3       & 84     & 0.78±0.09 & 1.00±0.00 & 1.00±0.00    & 1.00±0.00 & 1.00±0.00 & 1.00±0.00 & 1.00±0.00  \\
HIV FE         & 733    & 0.87±0.04 & 1.00±0.02 & 1.00±0.02    & 0.98±0.05 & 0.98±0.05 & 0.63±0.30 & 1.00±0.02  \\
HIV GSL3       & 786    & 0.86±0.04 & 0.99±0.02 & 0.99±0.02    & 0.98±0.06 & 0.99±0.02 & 0.80±0.20 & 0.99±0.02  \\
HIV PBS        & 188    & 0.92±0.02 & 1.00±0.01 & 1.00±0.01    & 1.00±0.02 & 0.99±0.03 & 0.90±0.11 & 1.00±0.01  \\
Intron gpII    & 181    & 0.46±0.13 & 0.65±0.16 & 0.66±0.17    & 0.63±0.17 & 0.50±0.28 & 0.49±0.18 & 0.65±0.15  \\
IRES HCV       & 764    & 0.65±0.11 & 0.89±0.16 & 0.77±0.31    & 0.86±0.17 & 0.69±0.38 & 0.85±0.08 & 0.89±0.08  \\
IRES Picorna   & 181    & 0.84±0.07 & 0.97±0.03 & 0.95±0.06    & 0.96±0.04 & 0.95±0.04 & 0.84±0.11 & 0.96±0.04  \\
K chan RES     & 124    & 0.74±0.10 & 0.99±0.02 & 0.98±0.05    & 0.89±0.19 & 0.95±0.08 & 0.57±0.26 & 0.95±0.12  \\
Lysine         & 80     & 0.50±0.13 & 0.73±0.13 & 0.70±0.19    & 0.72±0.18 & 0.66±0.16 & 0.49±0.16 & 0.72±0.15  \\
Retroviral psi & 89     & 0.88±0.03 & 0.93±0.03 & 0.94±0.03    & 0.94±0.03 & 0.93±0.04 & 0.73±0.13 & 0.93±0.04  \\
S box          & 91     & 0.60±0.10 & 0.75±0.12 & 0.77±0.16    & 0.79±0.14 & 0.67±0.24 & 0.53±0.16 & 0.77±0.12  \\
SECIS          & 114    & 0.44±0.16 & 0.59±0.21 & 0.63±0.21    & 0.58±0.25 & 0.54±0.25 & 0.38±0.24 & 0.62±0.20  \\
sno 14q I II   & 44     & 0.75±0.10 & 0.93±0.10 & 0.89±0.16    & 0.85±0.20 & 0.89±0.19 & 0.57±0.27 & 0.91±0.13  \\
SRP bact       & 114    & 0.48±0.16 & 0.66±0.21 & 0.66±0.20    & 0.64±0.24 & 0.65±0.21 & 0.51±0.21 & 0.62±0.25  \\
SRP euk arch   & 122    & 0.51±0.20 & 0.63±0.29 & 0.63±0.29    & 0.65±0.28 & 0.65±0.25 & 0.50±0.25 & 0.62±0.28  \\
T-box          & 18     & 0.68±0.15 & 0.78±0.17 & 0.75±0.25    & 0.67±0.24 & 0.70±0.17 & 0.59±0.20 & 0.74±0.15  \\
TAR            & 286    & 0.87±0.04 & 0.99±0.03 & 0.99±0.02    & 0.99±0.03 & 0.98±0.04 & 0.83±0.20 & 0.99±0.04  \\
THI            & 321    & 0.45±0.10 & 0.69±0.15 & 0.68±0.19    & 0.69±0.17 & 0.51±0.29 & 0.48±0.18 & 0.66±0.20  \\
tRNA           & 2039   & 0.43±0.12 & 0.75±0.21 & 0.85±0.16    & 0.82±0.19 & 0.76±0.27 & 0.65±0.23 & 0.72±0.22  \\
U1             & 82     & 0.63±0.17 & 0.80±0.17 & 0.83±0.14    & 0.79±0.18 & 0.81±0.14 & 0.67±0.20 & 0.77±0.17  \\
U2             & 112    & 0.64±0.16 & 0.76±0.17 & 0.83±0.17    & 0.77±0.19 & 0.73±0.22 & 0.59±0.19 & 0.75±0.18  \\
U6             & 30     & 0.83±0.06 & 0.93±0.05 & 0.89±0.09    & 0.90±0.08 & 0.88±0.10 & 0.71±0.14 & 0.93±0.06  \\
UnaL2          & 138    & 0.77±0.08 & 0.93±0.08 & 0.92±0.09    & 0.89±0.15 & 0.91±0.10 & 0.64±0.29 & 0.94±0.08  \\
yybP-ykoY      & 127    & 0.39±0.14 & 0.58±0.20 & 0.59±0.24    & 0.58±0.25 & 0.40±0.33 & 0.46±0.21 & 0.56±0.20  \\	
\hline	
\hline	
Pooled Average & 249.33 & 0.63      & 0.84      & 0.86         & 0.85      & 0.8       & 0.67      & 0.83 \\
\hline	
\end{tabular}
\end{tiny}
\caption{Average positive predictive value (PPV) scores ($\pm$ one standard deviation) for
{\em pairwise global} alignment of {\tt RNAmountAlign} and four widely used
RNA sequence/structure alignment algorithms on the benchmarking set of
8976 pairwise alignments from the
{\tt BRaliBase K2} database \cite{Gardner.nar05}.
For each indicated Rfam family, the
the number of alignments (NumAln), sequence identity (SeqId), and
PPV-scores  for
{\tt RNAmountAlign}, {\tt LocARNA}, {\tt LARA}, {\tt FOLDALIGN}, and
{\tt DYNALIGN} are listed, along with
Pooled averages over all 8976 pairwise
alignments.  Parameters used in Eq~(\ref{eqn:similarityMeasure})
of the main text
for {\tt RNAmountAlign} were similarity matrix RIBOSUM85-60, structural
similarity weight $\gamma=1/2$, gap initiation $g_i=-3$, gap
extension $g_e=-1$.
}
\label{table:globalPPV}
\end{table*}

\begin{figure*}
\centering
\includegraphics[width=0.8\textwidth]{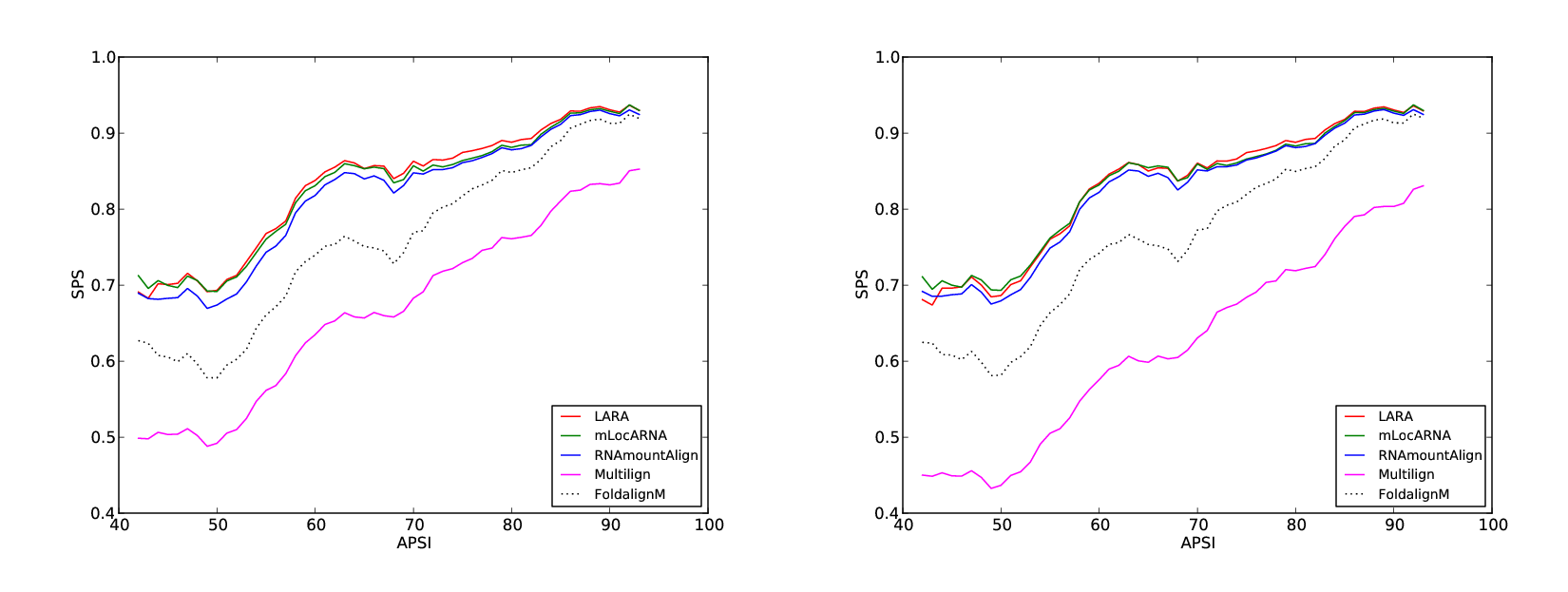}

\caption{Average pairwise sensitivity (left) and positive predictive value (right) for \textit{multiple global alignments} using 
{\tt RNAmountAlign}, {\tt LARA}, {\tt mLocARNA}, {\tt FoldalignM} and {\tt Multilign} in the
k5 {\tt BRAliBase 3} database used for benchmarking. Note that in our definition of $Sen$ and $PPV$, pairs of the form $(X,\mbox{\textemdash})$ and $(\mbox{\textemdash},X)$ are also counted while SPS is the average pairwise sensitivity only considering aligned residue pairs (Fig~\ref{fig:multiFSCI}). However, the results with and without gap counts, indicated in this Fig and Fig~\ref{fig:multiFSCI}, respectively, are very close.
}
\label{fig:multiSenPPV}
\end{figure*}

\begin{table*}[]
\begin{small}
\begin{tabular}{|c|c|c|c|c|c|c|c|c|c|}
\hline
$p_A$   & $p_C$   & $p_G$   & $p_U$   & $p_{\op}$ & $p_{\cp}$ & $p_{\bullet}$   & $std_{\op}$ & $std_{\cp}$ & $std_{\bullet}$   \\
\hline
0.00 & 0.00 & 0.00 & 1.00 & 0.000000 & 0.000000 & 1.000000 & 0.000000 & 0.000000 & 0.000000 \\
0.00 & 0.00 & 0.05 & 0.95 & 0.000533 & 0.000533 & 0.998933 & 0.000292 & 0.000292 & 0.000583 \\
0.00 & 0.00 & 0.10 & 0.90 & 0.001396 & 0.001396 & 0.997209 & 0.000818 & 0.000818 & 0.001636 \\
0.00 & 0.00 & 0.15 & 0.85 & 0.002704 & 0.002704 & 0.994592 & 0.001548 & 0.001548 & 0.003096 \\
0.00 & 0.00 & 0.20 & 0.80 & 0.004785 & 0.004785 & 0.990431 & 0.002863 & 0.002863 & 0.005725 \\
0.00 & 0.00 & 0.25 & 0.75 & 0.008039 & 0.008039 & 0.983922 & 0.004992 & 0.004992 & 0.009983 \\
0.00 & 0.00 & 0.30 & 0.70 & 0.013641 & 0.013641 & 0.972717 & 0.008488 & 0.008488 & 0.016976 \\
0.15 & 0.20 & 0.15 & 0.50 & 0.198666 & 0.198666 & 0.602668 & 0.031304 & 0.031304 & 0.062607 \\
0.15 & 0.20 & 0.20 & 0.45 & 0.244486 & 0.244486 & 0.511027 & 0.028368 & 0.028368 & 0.056737 \\
0.15 & 0.20 & 0.25 & 0.40 & 0.280658 & 0.280658 & 0.438684 & 0.023478 & 0.023478 & 0.046957 \\
0.15 & 0.20 & 0.30 & 0.35 & 0.306193 & 0.306193 & 0.387613 & 0.018226 & 0.018226 & 0.036452 \\
0.15 & 0.20 & 0.35 & 0.30 & 0.319277 & 0.319277 & 0.361446 & 0.014271 & 0.014271 & 0.028541 \\
0.15 & 0.20 & 0.40 & 0.25 & 0.320472 & 0.320472 & 0.359056 & 0.014868 & 0.014868 & 0.029735 \\
0.15 & 0.20 & 0.45 & 0.20 & 0.310048 & 0.310048 & 0.379905 & 0.018890 & 0.018890 & 0.037781 \\
0.15 & 0.20 & 0.50 & 0.15 & 0.289160 & 0.289160 & 0.421679 & 0.023603 & 0.023603 & 0.047205 \\
0.15 & 0.20 & 0.55 & 0.10 & 0.259201 & 0.259201 & 0.481598 & 0.027322 & 0.027322 & 0.054644 \\
0.15 & 0.20 & 0.60 & 0.05 & 0.223416 & 0.223416 & 0.553168 & 0.027906 & 0.027906 & 0.055813 \\
0.15 & 0.20 & 0.65 & 0.00 & 0.183844 & 0.183844 & 0.632311 & 0.026849 & 0.026849 & 0.053698 \\
0.15 & 0.25 & 0.00 & 0.60 & 0.009383 & 0.009383 & 0.981234 & 0.008960 & 0.008960 & 0.017920\\
\hline
\end{tabular}
\end{small}
\centering
\caption{Initial portion of a table that determines expected base pairing probabilities
$p_{\op},p_{\bullet},p_{\cp}$ as a function of nucleotide probabilities
$p_A,p_C,p_G,p_U$. The full table (not shown) has 1770 rows.
To determine average base pairing probabilities, given
nucleotide probabilities $p_A,p_C,p_G,p_U$, a total of $N=10000$ RNA sequences of length
$n=200$ were randomly generated to have the given expected nucleotide frequency. To compute
$p_{\op}$ [ resp. $std_{\op}$ ], a library call
of function {\tt pf\_fold()} from Vienna RNA Package \cite{Lorenz.amb11} was made in order to determine
$Prob[\mbox{$i$ pairs to right}] = \sum_{i=1}^n \sum_{j=i+1}^n p_{i,j}$ for position in each
sequence, and the average [ resp. standard deviation ] was taken over all sequences and values $i=1,\ldots,n$. In a similar fashion,
$p_{\bullet}$ and $p_{\cp}$ were determined.
}
\label{table:precomputedProb} 
\end{table*}

\begin{figure}
\centering
\includegraphics[width=0.3\textwidth]{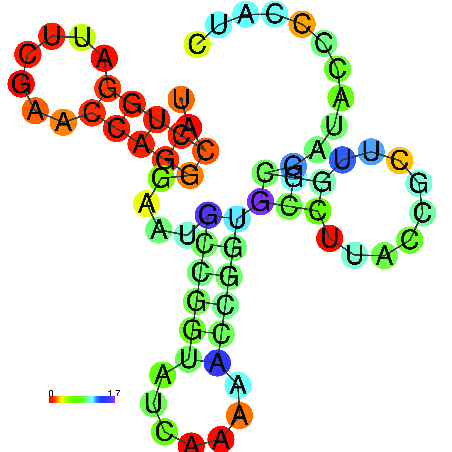}
\caption{Illustration of a potential weakness of {\tt RNAmountAlign}.
Using {\tt RNAmountAlign} genome-scanning software, semiglobal alignments of 
the query tRNA AB031215.1/9125-9195 were made with each 300 nt window
(successive window overlap of 200 nt) of the
{\em E. coli} str. K-12 substr. MG1655 genome. This figure shows the 
MFE structure, color-coded by positional entropy  \cite{Huynen.jmb97},
for the alignment
of positions 696097-696164 with score $-7.70$, $p$-value of
$4.145010 \cdot 10^{-6}$.
(gap costs $g_i =-3$, $g_i=-1$, $\gamma=0.5$,
scaling factor $\alpha_{\mbox{\tiny seq}} = 0.447648$,
shift term $\alpha_{\mbox{\tiny str}} = 0.304766$,
$\gamma=1/2$).
However, this RNA is clearly not a tRNA, since
the three loops are not within the scope of a multiloop, and the variable
loop is located in the wrong position, and the large positional entropy
suggests that there is not an unambiguous structure. Moreover, 
this sequence is not one of the 40 tRNA genes/pseudogenes on the 
plus-strand predicted by {\tt tRNAscan-SE} \cite{Lowe.nar16}.
}
\label{fig:falsePosTRNA}
\end{figure}

\end{document}